\documentclass{aa}

\usepackage{hyperref}
\usepackage{txfonts}
\usepackage{natbib}

\bibpunct{(}{)}{;}{a}{}{,}

\hypersetup{
 colorlinks=true,
 allcolors=blue
}

\begin{document}

\date{Received 17 April 2024 / Accepted 5 November 2024}

\title{Evolutionary tracks of binary neutron star progenitors\\across cosmic times}

\titlerunning{Evolutionary tracks of binary neutron star progenitors across cosmic times}

\author{Cl\'ement~Pellouin\inst{\ref{iap}}$^,$\inst{\ref{tau}}
\and Irina~Dvorkin\inst{\ref{iap}}$^,$\inst{\ref{iuf}}
\and L\'eonard~Lehoucq\inst{\ref{iap}}}

\institute{Sorbonne Universit\'e, CNRS, UMR 7095, Institut d’Astrophysique de Paris (IAP), 98 bis boulevard Arago, 75014 Paris,
France\label{iap}; e-mail: \texttt{pellouin@iap.fr};
\and 
School of Physics and Astronomy, Tel Aviv University, Tel Aviv 6997801, Israel\label{tau};
\and
Institut Universitaire de France, Minist\`ere de l’Enseignement Sup\'erieur et de la Recherche, 1 rue Descartes, 75231 Paris Cedex F-05, France\label{iuf}
}

\authorrunning{C. Pellouin et al.}

\abstract{
Recent discoveries of gravitational wave sources have advanced our knowledge about the formation of compact object binaries. At present, many questions about the stellar origins of binary neutron stars remain open. We explore the evolution of binary neutron star progenitors with the population synthesis code COSMIC. We identify three dominant evolutionary tracks to form neutron star binaries that merge within the age of the Universe: a scenario that includes a common envelope phase between the first neutron star and its companion, a scenario with almost equal-mass progenitors that evolve quasi-simultaneously and which features a double-core common envelope, and a scenario involving the accretion-induced collapse of an oxygen-neon white dwarf into a neutron star. We show that the distribution of time delays between stellar formation and binary neutron star merger at a given progenitor metallicity does not follow a power-law, but instead features a complex structure that reflects the progenitor properties and the relative contribution of each evolutionary track. We also explore the evolution of the merger rate density with redshift and show that the scenario involving the accretion-induced collapse could be dominant at high redshifts. These results can have important implications for the study of the chemical enrichment of galaxies in r-process elements produced in kilonovae; and of short gamma-ray bursts offsets in their host galaxies.
}
  
\keywords{Stars: binaries -- Stars: neutron -- Stars: evolution -- Gravitational waves
            }

\maketitle


\section{Introduction}

The discovery of gravitational waves (GW) from merging binary compact objects \citep{2016PhRvL.116f1102A} has revolutionized observational astronomy. The observations made by the Advanced LIGO \citep{2015CQGra..32g4001L} and Advanced Virgo \citep{2015CQGra..32b4001A} detectors have led to major breakthroughs in the study of compact objects and their stellar progenitors, and opened new avenues for fundamental physics and cosmology \citep{2023PhRvX..13d1039A}. While the vast majority of the detected sources are binary black holes (BBH), two binary neutron star (BNS) mergers have been observed. The first detection (GW~170817, \citealt{2017PhRvL.119p1101A,2017ApJ...848L..12A}), was a multi-messenger source, having been observed in the optical \citep{2017Sci...358.1556C}, infrared, X-ray, gamma-ray \citep{2017ApJ...848L..14G, 2017ApJ...848L..15S} and radio \citep{2018Natur.561..355M, 2019Sci...363..968G}. These observations confirmed the long-standing hypothesis that some short gamma-ray bursts originate from a jet launched during the BNS merger and allowed to study the geometry of the jet. In addition, the observations in infrared, optical and ultraviolet confirmed the presence of a kilonova. This phenomenon consists in the thermal radiation from the radioactive decay of massive elements produced in the optically-thick BNS merger ejecta by rapid neutron capture (r-process). This confirmed that BNS mergers are sites of r-process nucleosynthesis (see e.g. \citealt{2017ApJ...851L..21V}). The second BNS detection (GW~190425), while lacking observed electromagnetic counterparts, is also remarkable in that the total mass of the system ($3.4~M_{\odot}$) is significantly larger \citep{2020ApJ...892L...3A} than any of the other known BNSs (Galactic pulsars as well as GW~170817). These detections also allowed to estimate the merger rate in the local Universe, albeit with a large uncertainty, as $10 - 1700~\mathrm{Gpc^{-3}\cdot yr^{-1}}$ \citep{2023PhRvX..13a1048A}. It is interesting to note that Galactic binary pulsars and BNS systems observed via their GW emission probe different evolutionary stages of the same underlying population: from large separation to small separation (due to GW emission) to merger. 

While the number of known BNS systems is currently relatively small, it is expected to grow in the very near future. Based on current population models, at design sensitivity the LIGO-Virgo-Kagra network is expected to observe up to about 100 systems per year \citep{2023ApJ...958..158K}. Moreover, third generation GW detectors, Einstein Telescope and Cosmic Explorer, planned for the 2030s, are expected to detect up to $10^5$ BNS mergers per year \citep{2020JCAP...03..050M, 2021arXiv210909882E, 2023JCAP...07..068B}. 

The formation scenarios of binary compact objects, and in particular BNSs, were extensively studied using methods with different degrees of approximations. In particular, population synthesis codes \citep[e.g.][]{1996A&A...309..179P,2002MNRAS.329..897H,2002ApJ...572..407B,2015MNRAS.451.4086S,2016MNRAS.462.3302E,2017NatCo...814906S,2018MNRAS.474.2959G,2018MNRAS.481.1908K,2020ApJ...898...71B,2022ApJS..258...34R,2023ApJS..264...45F,2023MNRAS.524..426I} became a standard tool in compact binary studies thanks to their computational efficiency, and have been used to estimate the BNS merger rates (see references in \citealt{2022LRR....25....1M}). Most of these codes follow stellar evolution using detailed simulations for \emph{single} stars, assuming these are not affected by the presence of the companion. BPASS \citep{2016MNRAS.462.3302E} and POSYDON \citep{2023ApJS..264...45F}, on the contrary, rely on detailed binary evolution simulations. Stellar properties at each time step are then computed either using fitting formulae (e.g. COSMIC, \citealt{2020ApJ...898...71B}; COMPAS, \citealt{2022ApJS..258...34R}; StarTrack, \citealt{2002ApJ...572..407B}) or interpolation on pre-computed grids (e.g. SEVN, \citealt{2023MNRAS.524..426I}; POSYDON). In population synthesis codes based on single-star evolution, the processes related to binarity (e.g. mass exchanges, orbital evolution) are computed using phenomenological prescriptions. As most BNSs are likely to be formed by isolated binaries, this method is particularly useful, since it allows to model cosmological populations of compact binaries and to probe large portions of the parameter space related to binary evolution. In addition, several codes have been developed for the study of \emph{dynamical} effects in dense stellar systems \citep[e.g.][]{2013MNRAS.431.2184G,2020ApJS..247...48K}. While the details of different population synthesis codes vary, several general conclusions on BNS populations emerge. 

First of all, many recent studies find that most BNSs are expected to form from isolated binary massive stars, rather than through dynamical interactions in dense environments \citep{2018A&A...615A..91B,2019ApJ...886....4Z,2020ApJ...888L..10Y,2020ApJ...901L..16F}. Indeed, while three- and four-body interactions can drive BBHs to merge in dense stellar clusters, this is not the case for BNSs. The reason for this difference is that BBHs, being more massive, dominate the cores of stellar clusters, thereby preventing mass segregation of neutron stars (NSs). This effect greatly reduces the rate of dynamical interactions NSs can undergo. Additionally, because NSs are less massive, they are more likely to be ejected out of the clusters upon formation due to their natal kicks.

The second conclusion is that within the isolated formation channel, a specific evolutionary track is favored. It features a binary massive star at Zero Age Main Sequence (ZAMS),  a phase of common envelope (CE) between a giant and a NS, and a later phase of mass transfer between the stripped giant star and the NS \citep[\textit{case BB} mass transfer, e.g.][]{2015MNRAS.451.2123T}. Both mass transfer episodes shrink the orbit considerably, so that the resulting binary compact object is more likely to merge within the Hubble time, $t_\mathrm{Hubble} = 13.7~\mathrm{Gyr}$. Moreover, the secondary explodes as an ultra-stripped supernova (SN), which leads to a smaller natal kick compared to a standard core-collapse SN, thereby increasing the probability that the system remains gravitationally bound after the second supernova.

More generally, several studies have pointed out that the natal kicks of core-collapse SNe that form NSs should be weaker than those that form BHs, of the order of $20~\mathrm{km}\cdot\mathrm{s}^{-1}$ \citep[e.g.][]{2018MNRAS.480.2011G}. Conversely, BH natal kicks are typically one order of magnitude larger \citep{2005MNRAS.360..974H}, though some more recent models scale them down by a factor $1.4M_\odot / M_\mathrm{BH}$ so that NSs and BHs receive the same momentum after the collapse \citep[e.g.][]{2022ApJS..258...34R}. Fallback accretion may further reduce the BH natal kick \citep{2012ApJ...749...91F}. Low NS kicks are naturally expected in the case of ultra-stripped SNe, electron capture SNe and accretion-induced collapse (AIC) of a white dwarf (WD) into a NS. Such low kicks in AIC events may increase the survival rate of BNS progenitors \citep{2018MNRAS.474.2937C}. As a result, AIC could play an important role in the formation of BNSs, and we briefly summarize this process here.

AIC occurs when a heavy (O/Ne/Mg or Si/O) white dwarf accretes matter from its stellar companion with a slow enough rate ($\lesssim 10^{-9}M_{\odot}\cdot\mathrm{yr}^{-1}$), leading the WD to collapse and form a NS. In general, AIC can take place either in a double-degenerate case (WD-WD merger) or a single-degenerate case (WD accreting from a companion star that overflows its Roche lobe). \citet{2019MNRAS.484..698R} suggested that double-degenerate AIC events operating in WD binary systems could be an important step in forming merging BNSs observed with ground-based GW interferometers. In this work, we will show the possibility that single-degenerate AIC events can also contribute to the population of merging BNSs at high redshifts. AIC events are expected to produce faint optical transients, but these would be short-lived (lasting from a few days to a week) and under-luminous with respect to regular SNIa \citep{2023MNRAS.525.6359L}. Such events are therefore difficult to observe, and there have been no detections reported to date, although some indirect evidence exists \citep[see e.g.][for a review]{2020RAA....20..135W}.

Previous studies of BNS populations have also explored the time delays $\Delta t_\mathrm{delay}$ between the formation of the stellar binary (ZAMS) and the merger of the BNS. This delay is defined as the sum of the stellar evolution part (the time for the massive stellar binary to form a compact object binary), $\Delta t_\mathrm{evol}$; and the orbital evolution due to the emission of GWs (orbital decay of a compact object up to coalescence), $\Delta t_\mathrm{GW}$. Usually, a single distribution of the time delays is assumed  for an entire cosmological population, and it is found that the distribution follows closely a power-law, $P\left(\Delta t_\mathrm{delay}\right) \propto \Delta t_\mathrm{delay}^{-1}$ \citep[e.g.][]{2018MNRAS.474.2937C,2019MNRAS.490.3740N}. In fact, this is the expected distribution if the delays are dominated by the GW phase (i.e. assuming negligible stellar evolution time), and if the distribution of initial semi-major axes $a_i$ is log-uniform, since the GW delay scales as $\Delta t_\mathrm{GW}\propto a_i^4$. However the question arises as to whether this distribution is universal, i.e. whether it applies to all the evolutionary tracks of BNS progenitors at all epochs; or if it depends on the stellar evolution parameters.

In this work we explore the evolutionary tracks of BNS progenitors using the population synthesis code COSMIC \citep{2020ApJ...898...71B} and study in detail the distribution of delay times through cosmic history. Throughout the paper we focus on systems that merge within the age of the Universe. Our population synthesis parameter choices, metallicity sampling protocol, as well as the procedure to identify the evolutionary tracks and the merger rate density calculation are explained in Sect.~\ref{sec:pop-synth}. With our assumptions on stellar evolution parameters, we identify three evolutionary tracks: a \emph{standard} one, involving \textit{case BB} mass transfer; a track involving two stars with almost identical ZAMS masses which leads to synchronized evolution and a phase of CE with both stars in their giant phase; and a track that involves the AIC of one of the progenitors. These tracks are discussed in Sect.~\ref{sec:results_evol_tracks}. We then focus on the dependence on metallicity and explore in detail the properties of the progenitor stellar binaries that contribute to each track in Sect.~\ref{sec:fixed_met_progenitors}. We discuss the properties of the BNS populations in each track in Sect.~\ref{sec:bns-pops}. We find that the time delay distributions at a given metallicity do not follow a simple power-law, but instead present a complex structure that depends on the evolutionary track. Having identified the dominant evolutionary tracks at each metallicity, we explore the properties of the resulting cosmological population in Sect.~\ref{sec:merger-rate}. In particular, we find that the \emph{AIC} evolutionary track has very short delay times $\Delta t_\mathrm{GW}$ and is dominant at high redshift. However at low redshift, the equal-mass evolutionary track is dominant, and it is the one expected to contribute to the population observed with GW detectors. We discuss some interesting applications and consequences of our results as well as some caveats to our analysis in Sect.~\ref{sec:discussion}.

\section{Modelling binary neutron star populations}
\label{sec:pop-synth}

\subsection{Population synthesis code}

In this work, we use the COSMIC population synthesis code \citep{2020ApJ...898...71B} to simulate a population of BNSs. COSMIC implements stellar evolution using SSE \citep{2000MNRAS.315..543H} and binary interactions using BSE \citep{2002MNRAS.329..897H}, with several modifications to account for recent updates to binary evolution. We use the default parameters of version 3.4.0, with a few exceptions that are discussed in this section. The full parameter list used in this study can be found in Appendix~\ref{ap:cosmic_params}.

We note that the goal of this study is not to perform a systematic sampling of the possible parameter values; but rather to focus on one physically-motivated model and focus on the evolutionary tracks that lead to the formation of BNSs, that we describe in Sect.~\ref{sec:results_evol_tracks}.

We assume standard properties of the binaries at ZAMS. The masses of the primary follow a broken power-law distribution \citep{2001MNRAS.322..231K}, and mass ratios are uniformly distributed \citep[e.g.][]{2012Sci...337..444S}. We follow the distributions of initial semi-major axes and eccentricities introduced by \citet{2012Sci...337..444S}.

Stellar winds are treated using recent corrections to BSE discussed in \citealt{2020ApJ...898...71B} (corresponding to \texttt{windflag}~$=3$). In this study, we do not introduce Eddington-limited winds as proposed by \citet{2008A&A...482..945G}, where stronger winds are less affected by metallicity (see also \citealt{2018MNRAS.474.2959G}). The average accretion rate per orbit onto the companion is assumed to follow the \citet{1944MNRAS.104..273B} mechanism and is therefore limited. In particular, most mass transfer events are non-conservative. Thus, any event of AIC is only possible following a phase of stable mass transfer onto a WD companion. Stellar winds mostly impact binary evolution because of the single-star loss of mass. Wind velocities are assumed to depend on the stellar type \citep{2008ApJS..174..223B}.

Due to the envelope expansion in the late stages of stellar evolution, the stellar radius can exceed its Roche-lobe radius ($R_\mathrm{L}$), which is always computed at periastron and assuming the expression from \citet{1983ApJ...268..368E}. When mass transfer starts, the orbit is assumed to instantaneously circularize at periastron. Mass transfer can be stable (Roche lobe overflow) or unstable, in which case this leads to a CE (see e.g. the review by \citealt{2013A&ARv..21...59I}). The treatment of the boundary between stable and unstable mass transfer rates depends on the stellar type of (only) the donor star and on the mass ratio $q = M_\mathrm{don} / M_\mathrm{comp}$. If $q > q_\mathrm{crit}$, mass transfer is dynamically unstable and a phase of CE follows. We use the values for $q_\mathrm{crit}$ proposed by \citet{2019MNRAS.490.3740N} and also used in e.g. COMPAS (\texttt{qcflag} $=5$). In particular, with these values, mass transfers from stripped donors are always dynamically stable.

In the isolated binary formation scenario, the CE phase allows the orbital separation to shrink enough for the compact object binary formed after the stellar explosions to merge within $t_\mathrm{Hubble}$. COSMIC treats CE in a simple parametric way using the \textit{alpha lambda} formalism \citep{1984ApJ...277..355W}, where $\alpha_\mathrm{CE}$ describes the fraction of the orbital energy that is transferred to the envelope, leading to its expansion and possible ejection; and $\lambda$ (of order unity) describes the stiffness of the envelope density profile. In this study, we fix $\alpha_\mathrm{CE} = 1$ (we do not include external energy terms as discussed in \citealt{2019ApJ...883L..45F, 2021MNRAS.502.4877S}); and a variable $\lambda$ that depends on the stellar type (see \citealt{2014A&A...563A..83C}). We assume that stellar companions without a clear core-envelope boundary automatically lead to a merger during the CE phase (\texttt{cemergeflag} $=1$), as introduced by \citet{2008ApJS..174..223B}. We do not deduce the final core mass of the donor star in the case of \textit{case BB} mass transfer using the expressions for the rates from \citet{2015MNRAS.451.2123T}, and instead assume the expression from \citet{2002MNRAS.329..897H} (\texttt{cehestarflag} $=0$).

Natal kicks are given to the compact objects that form following SNe, mostly because of asymmetries in the ejected material during the explosion (see e.g. \citealt{1994A&A...290..496J, 2015A&A...577A..48W}). For core-collapse SNe, we assume that the natal kicks follow a Maxwellian distribution with a kick velocity dispersion $\sigma_\mathrm{k} = 265~\mathrm{km}\cdot\mathrm{s}^{-1}$ \citep{2005MNRAS.360..974H}. For electron-capture SNe, ultra-stripped SNe and accretion-induced collapse (AIC) events, we assume that the kicks are reduced and have a dispersion $\sigma_\mathrm{k, low} = 20~\mathrm{km}\cdot\mathrm{s}^{-1}$. This parameter is especially important in this study as most evolutionary tracks involve one or several of these events with lower kicks, and therefore higher probabilities for the binary to survive the explosions. We do not assume that the kick velocity is affected by the ejected mass or the remnant mass as e.g. \citet{2020ApJ...891..141G} (\texttt{kickflag} $=0$). We infer the remnant mass following the \textit{rapid} mechanism for the SN explosion \citep{2012ApJ...749...91F}, with updates from \citet{2020ApJ...891..141G} (\texttt{remnantflag} $=3$). This leads to a mass gap between NSs and BHs, contrary to the \textit{delayed} mechanism.

Finally, the Solar metallicity is set to $Z_\odot = 0.014$ \citep{2009ARA&A..47..481A}.

\subsection{Sampling protocol}

We create a grid of metallicities ranging from $Z = 9.5 \times 10^{-5}$ ($6.8\times 10^{-3}~Z_\odot$, the minimum metallicity allowed in BSE \citep{2002MNRAS.329..897H} and thus in COSMIC) to $Z=0.028$ ($2~Z_\odot$). In practice, this last metallicity bin is not used in our calculation of the merger rate, and the details of stellar evolution at super-solar metallicities are more uncertain. The metallicity grid we use is thus $0.000095$; $0.00014$; $0.00021$; $0.0003$; $0.00044$; $0.00065$; $0.00095$; $0.0014$; $0.0021$; $0.003$; $0.0044$; $0.0065$; $0.0095$; $0.014$(; $0.028$), i.e. $14$ ($15$) metallicity values. 

For each of these metallicities, we sample and evolve $9.55 \times 10^9$ binaries. We do not use the \texttt{match} feature of COSMIC that automatically stops the sampling once the properties of the BNS masses and/or semi-major axes and/or eccentricities have converged to a stable distribution (for a more complete description, see \citealt{2020ApJ...898...71B}). In our case, we decide to use a single random seed across all metallicities, meaning that the initial population is the same for all metallicities. For a given binary, the stochastic processes such as SN kick intensity and orientations are also seeded, which means that a direct comparison between individual binary systems across the metallicity bins is possible. A potential downside of our approach is that in some metallicity bins, the total number of BNS systems may be quite low ($\sim 10^4$) and thus the studied distributions slightly biased by the random seed. For a more accurate simulation, the random seed should also be varied across metallicities, but this is not expected to impact the results provided the BNS sample is large enough.

We focus on all binaries that produce BNS systems that remain gravitationally bound. Some of them have initial orbital properties (semi-major axis and eccentricity) that prevent them from merging within the Hubble time. We mark them as non-merging systems and remove them from our study.

\subsection{Classification of binary evolutionary tracks}
\label{sec:tracks}

In order to define evolutionary tracks, we design an automated classification scheme based on the output of COSMIC. We note that analogous procedures were employed by several authors in the context of BBH and NSBH systems (see e.g. \citealt{2018MNRAS.481.4009V, 2019MNRAS.490.3740N, 2021A&A...647A.153B, 2021MNRAS.508.5028B, 2022MNRAS.516.5737B, 2023MNRAS.524..426I}). In our case, for each simulation at a given metallicity, we store the information on the binaries that produce gravitationally-bound BNS systems, including all the stellar properties and binary properties at key stages of stellar evolution.

\begin{table}
    \caption{Stellar types indicators used in COSMIC.}
    \centering
    \resizebox{\hsize}{!}{
    \begin{tabular}{c|c}
    \hline\hline
        \textbf{Indicator} &  \textbf{Stellar Type}\\ \hline
        0 & Low-Mass Main Sequence (MS) star ($M < 0.7 M_\odot$)\\
        1 & MS star ($M > 0.7 M_\odot$)\\
        2 & Hertzsprung Gap (HG)\\
        3 & First Giant Branch (FGB)\\
        4 & Core Helium Burning (CHeB)\\
        5 & Early Asymptotic Giant Branch (EAGB)\\
        6 & Thermally Pulsating Asymptotic Giant Branch (TPAGB)\\
        7 & Naked Helium Star, Main Sequence (NHeMS)\\
        8 & Naked Helium Star, Hertzsprung Gap (NHeHG)\\
        9 & Naked Helium Star, Giant Branch (NHeGB)\\
        10 & Helium White Dwarf (WD)\\
        11 & Carbon/Oxygen WD\\
        12 & Oxygen/Neon WD\\
        13 & Neutron Star (NS)\\
        14 & Black Hole (BH)\\
        15 & Massless Remnant\\ \hline
    \end{tabular}
    }
    \tablefoot{The indicators correspond to the definitions of \citet{1998MNRAS.298..525P}.}
    \label{tab:stellar_types}
\end{table}

\begin{figure}
    \centering
    \resizebox{\hsize}{!}{\includegraphics{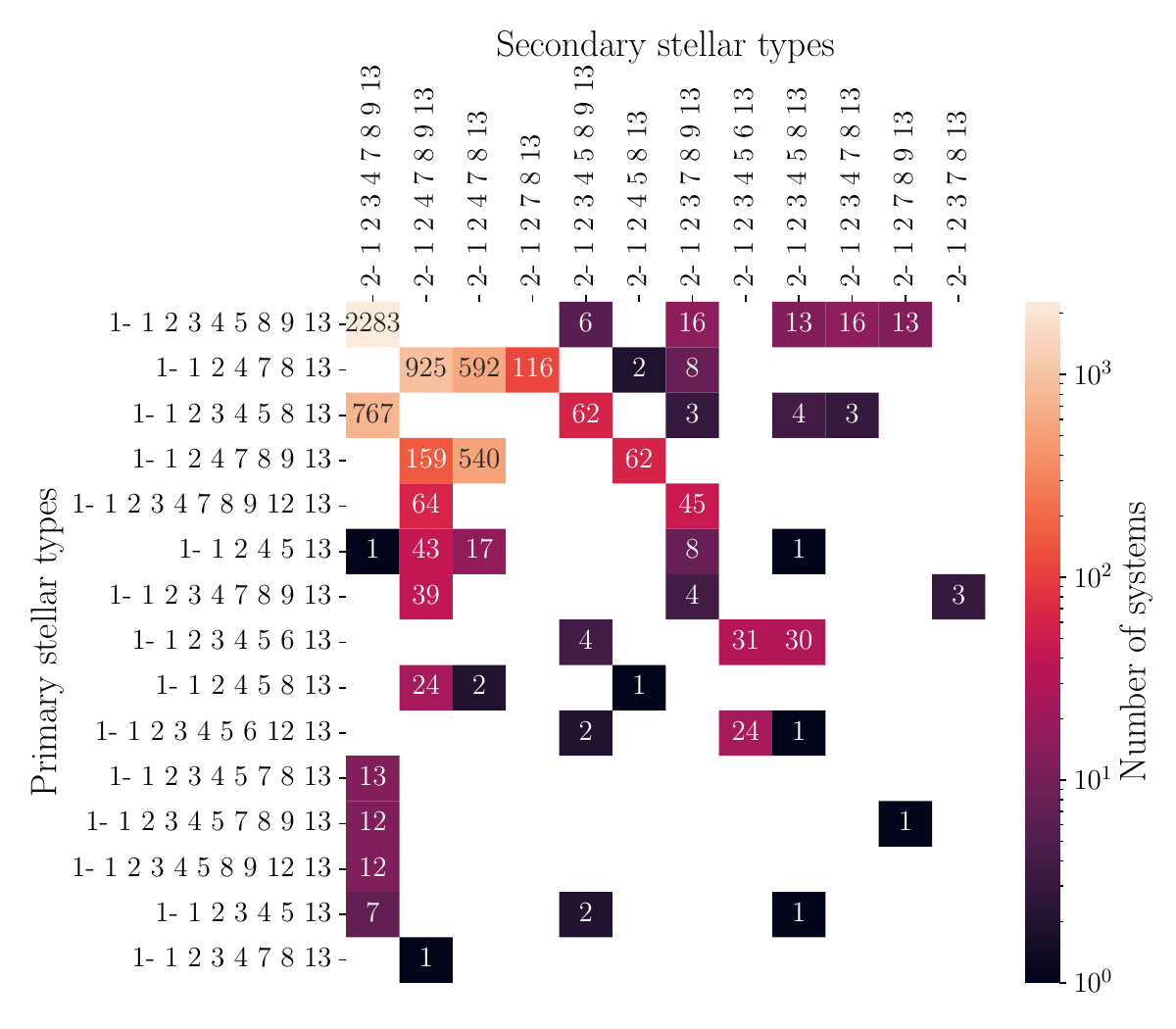}}
    \caption{Illustration of our classification scheme. Shown is the number of BNS systems for each sequence of stellar types of the primary (rows) and of the secondary (columns), at $Z = Z_\odot = 0.014$. The combination of primary and secondary evolutionary sequences creates $46$ evolutionary sequences, but $95\%$ of the systems are found in $12$ of them. The sequences of numbers on the y axis and on the x axis are the series of stellar types (see Table~\ref{tab:stellar_types}) for the primary and the secondary, respectively. Only systems with $\Delta t_\mathrm{delay} < t_\mathrm{Hubble}$ are shown.}
    \label{fig:heatmap_delay_filtered}
\end{figure}

The first stage is to identify evolutionary \textit{sequences}, which we define as the sequence of different stellar types for both the primary and the secondary (see Table~\ref{tab:stellar_types} for the list of stellar types used in COSMIC). An illustration of this procedure is presented in Fig.~\ref{fig:heatmap_delay_filtered}, calculated at solar metallicity. The vertical (primary stellar type) and horizontal (secondary stellar type) axes refer to evolutionary sequences of the primary and secondary, respectively. Each line (column) for the primary (secondary) corresponds to a particular sequence of stellar types as the evolution of the binary proceeds. For example, the sequence \texttt{1- 1 2 3 4 5 8 9 13} corresponds to primary stars which evolve off the main sequence (\texttt{1-2-3}) until they start burning helium in their core (\texttt{4}). After a brief time on the asymptotic giant branch (\texttt{5}), their helium core is stripped of its envelope and the naked He star evolves (\texttt{8-9}) until its explosion, forming a NS (\texttt{13}). All primary evolutionary sequences containing \texttt{12} correspond to systems forming through AIC. The intersections (in color) mark the combinations that allow for a formation of at least a BNS that merges within the age of the Universe. The color coding shows the number of systems in each evolutionary sequence. While the total number of sequences that leads to a BNS is quite large, only a few of them are sufficiently frequent. In the example shown here, $12$ most frequent evolutionary sequences make up to $95\%$ of the population of merging BNSs. 

This method does not guarantee \textit{a priori} that systems with the same evolutionary sequence have similar binary evolution history, since we did not account for binary interactions (such as CE), nor the order of specific events (such as SNe) in this classification. However, we verified that for all the evolutionary sequences defined with this method, the details of binary stellar evolution for all the binaries are indeed very similar. 

The dominant evolutionary sequences are not necessarily the same at all metallicities. We therefore repeat the same process over the populations at different metallicities to find all the dominant sequences. Overall, we extracted the $13$ evolutionary sequences that make up most of the population of BNSs across the different metallicity bins, and analysed each of them individually.  We found that these $13$ evolutionary tracks can in fact be grouped in only $3$ categories with different binary histories, that we describe in more detail in the Sects.~\ref{sec:standard_evol_track}~--~\ref{sec:aic_evol_track} below. In practice, we therefore leave a small fraction ($\sim 5\%$) of systems unclassified, but this doesn't affect the results presented here.

\subsection{Merger rate density}\label{sec:merger_rate_density}

We use the procedure described in \citet{2023MNRAS.526.4378L} to combine the populations from all the metallicity bins and simulate a realistic astrophysical population. Here we summarize briefly the main ingredients of the calculation.

Redshift and time are related by the relation
\begin{equation}
\label{eq:tz}
    \frac{dt}{dz} = \frac{H_0^{-1}}{(1+z)\sqrt{\Omega_\mathrm{m} (1+z)^3 + \Omega_\Lambda}}\,,
\end{equation}
where $H_0 = 68~\mathrm{km \cdot s^{-1} \cdot Mpc^{-1}}$ is the Hubble constant, $\Omega_\Lambda = 0.69$ is the fraction of dark energy in the total Universe energy density and $\Omega_\mathrm{m} = 0.31$ is the fraction of energy in matter (dark matter and baryonic matter). We also define the useful quantities $\Omega_\mathrm{b} = 0.045$, the baryonic fraction of energy density and $h_0 = 0.68$, the reduced Hubble constant (dimensionless). We assume here the cosmological values of \citet{2020A&A...641A...6P}.

We use the following functional form for the star formation rate (SFR) from \citet{2003MNRAS.339..312S}:
\begin{equation}
    \label{eq:sfr}
    \psi(z) = \nu \frac{a\,\exp{[b(z-z_\mathrm{m})]}}{a-b+b\,\exp{[a(z-z_\mathrm{m})]}}\,.
\end{equation}
We set the values of the functional parameters following \citet{2015MNRAS.447.2575V}: $\nu =  0.178\, M_\odot \cdot \mathrm{Mpc}^{-3} \cdot \mathrm{yr^{-1}}$, the amplitude of the peak of the SFR; $z_\mathrm{m} = 2$, the redshift of the peak of SFR and $a = 2.37$, $b = 1.80$, where $a$ is connected to the star formation rate density slope at $z < z_\mathrm{m}$ and $b-a$ to the slope for $z > z_\mathrm{m}$. We assume that the fraction of stars in binary systems is $50\%$ \citep{2012Sci...337..444S}.

While our simulations are performed in metallicity bins, the SFR is given as a function of redshift. In order to calculate the mean metallicity at any given redshift, we use the expression proposed by \citet{2016Natur.534..512B}:
\begin{equation}
\label{eq:metalmean}
    \overline{Z}(z) = \frac{y(1-R)}{\rho_b} \int_z^{z_\mathrm{max}} \frac{10^{0.5}\,\psi(z')}{H_0 (1+z')\sqrt{\Omega_\mathrm{m}(1+z')^3+\Omega_\Lambda}} \mathrm{d}z'\,,
\end{equation}
where $R = 0.27$ is the fraction of stellar mass ejected back into the interstellar medium; $y = 0.019$ is the mass ratio between the new metals created and total stellar mass; and $\rho_\mathrm{b} = 2.77 \cdot 10^{11} \Omega_\mathrm{b} h_0~ M_\odot \cdot \mathrm{Mpc}^{-3}$ is the baryon density. 

At a given redshift, the metallicity content may have a very high dispersion, due to differences between galaxies and non-homogeneous mixing within galaxies (note that we do not use the mass-metallicity relation as we do not model galaxies by mass). This dispersion influences the environments of stellar formation. For simplicity, we follow \citet{2021MNRAS.502.4877S} and adopt a log-normal distribution of metallicities around the average metallicity at a given redshift:
\begin{equation}
    \label{eq:metaldistrib}
    P(Z|z) = \frac{1}{\sqrt{2\pi\sigma^2}} \,\,\exp{\left(-\frac{(\log{(Z/Z_\odot)} - \log{(\overline{Z}(z)/Z_\odot)})^2}{2\sigma^2}\right)}\,,
\end{equation}
with $\sigma = 0.2$.

Finally, we can compute the BNS merger rate:
\begin{equation}
\label{eq:merger_rate}
\begin{split}
    \mathcal{R}_\mathrm{merg}(t) =  \int\displaylimits^{Z_\mathrm{max}}_{Z_\mathrm{min}} \int\displaylimits^{t_\mathrm{delay,max}}_{t_\mathrm{delay,min}} \left[ \right. & \alpha(Z)\,  \psi(t-\Delta t_\mathrm{delay})\,  P(\Delta t_\mathrm{delay}|Z) \, \\
    & \left. P(Z|t-\Delta t_\mathrm{delay})\,\mathrm{d}\Delta t_\mathrm{delay}\,\mathrm{d}Z \right]\, ,
\end{split}
\end{equation}
where the time delay distribution $P(\Delta t_\mathrm{delay}|Z)$ and the fraction of successful BNSs is taken directly from our COSMIC simulations.

\section{Results}\label{sec:results}

We show our results below, starting with the three evolutionary tracks we identified using representative systems as examples in Sect.~\ref{sec:results_evol_tracks}. We then show in more detail the properties of the progenitor (Sect.~\ref{sec:fixed_met_progenitors}) and BNS populations (Sect.~\ref{sec:bns-pops}) for each track, at constant metallicity and across metallicity bins. Finally we use the simulated binaries to obtain a cosmological population and show the resulting merger rate density (Sect.~\ref{sec:merger-rate}).

\subsection{Evolutionary tracks}\label{sec:results_evol_tracks}

\subsubsection{Unequal mass ratios: Standard channel}
\label{sec:standard_evol_track}

\begin{figure}
    \centering
    \resizebox{\hsize}{!}{\includegraphics{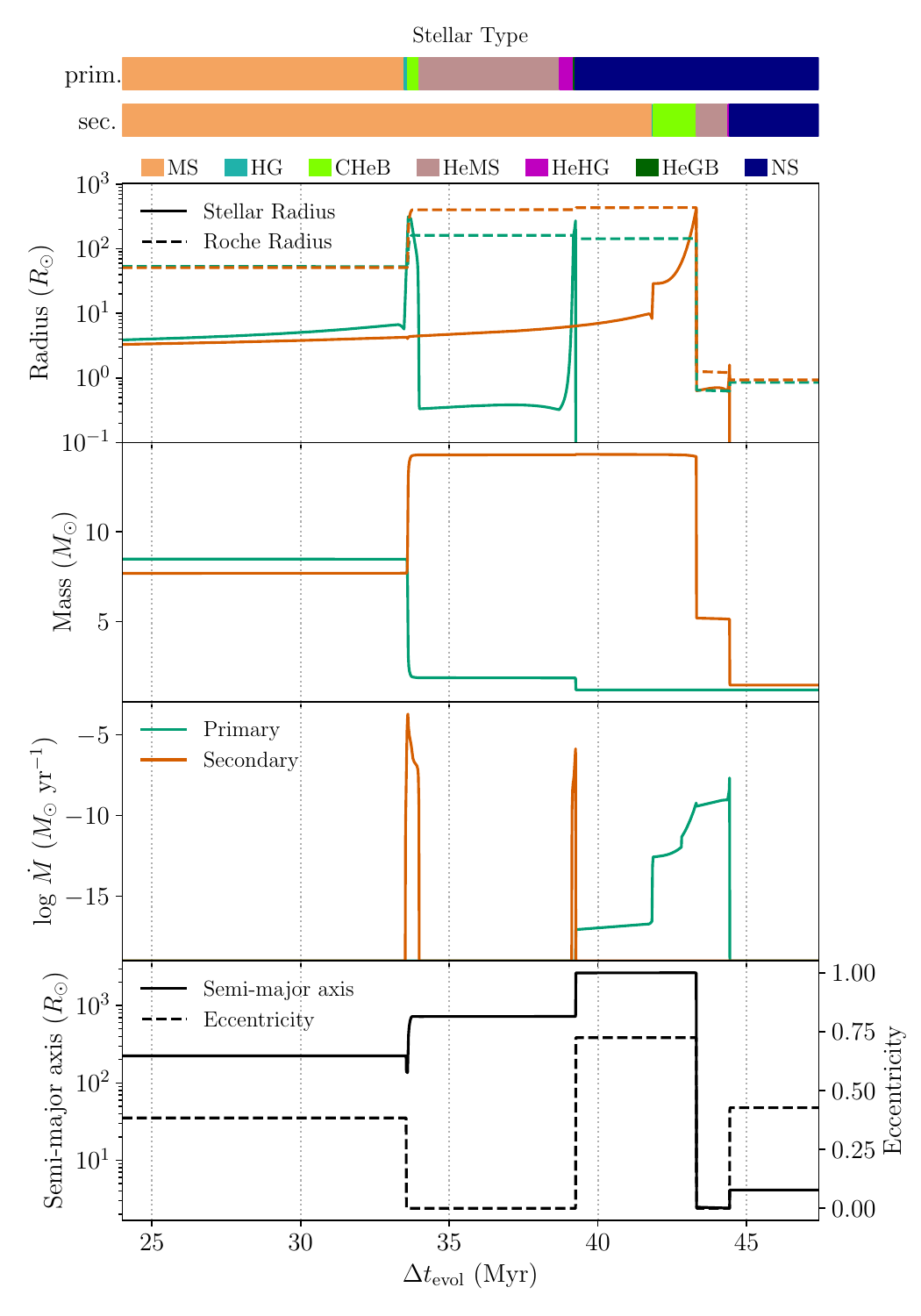}}
    \caption{Evolution of a binary system of the \emph{standard} evolutionary track forming a BNS at $Z = 0.00044$. The first $24~\mathrm{Myr}$ of main sequence evolution are hidden. The top color bars show the evolution of stellar types of the primary (top) and the secondary (bottom) during the binary evolution. Then, from top to bottom, we show the evolution of (1) the stellar radii (solid lines) and Roche radii (dashed lines) and (2) the masses of the two stars; (3) the mass growth rates and (4) the semi-major axis (solid line) and the eccentricity (dashed line) of the system. Quantities are shown in green for the primary and in orange for the secondary. When either of the stellar radii exceeds the Roche radius (panel (1)), a simultaneous accretion event occurs on the companion, in opposite colors (panel (3)). Accretion is also possible onto compact objects from stellar winds from the companion. Note that on panel (3), we show the stellar mass growth rate which accounts for both mass accretion and mass loss simultaneously. When $\dot{M}$ is not visible, this means that the star loses mass overall. When $\dot{M} > 0$ and $R < R_\mathrm{L}$, we see accretion of material onto a compact object (NS or WD) from the stellar winds of the companion.}
    \label{fig:standard_evol_example}
\end{figure}

The first evolutionary track we discuss corresponds to Channel I in \citet{2018MNRAS.481.4009V, 2019MNRAS.490.3740N, 2023MNRAS.524..426I}; Channel B in \citet[see appendix C]{2018MNRAS.481.1908K} and the dominant channel in \citet{2018MNRAS.474.2937C}, and was also already discussed in the literature as the dominant evolutionary track to produce NSBH systems (e.g. \citealt{2021MNRAS.508.5028B}; \citealt{2020ApJ...899L...1Z}) and BBHs \citep[e.g.][]{2018A&A...615A..91B}. 

We describe here a representative example of this evolutionary track as shown in Fig.~\ref{fig:standard_evol_example}. Initially, the binary has an unequal mass ratio (in this example, $q \simeq 0.9$ is quite on the high end of this distribution), such that the evolution timescale of the primary is shorter than that of the secondary. 

When the primary leaves the MS (after $\sim 33~\mathrm{Myr}$), it grows in radius and loses a significant amount of its mass ($\sim 6.5~M_\odot$) during a phase of stable mass transfer. This mass is transferred to the secondary. This phase lasts for $\lesssim 1~\mathrm{Myr}$. The primary is left without envelope as a NHeMS, until it again evolves into a NHeHG after $\sim 5~\mathrm{Myr}$. At that stage, a brief episode of mass transfer is triggered and the primary explodes in a SN. Given the progenitor mass at the moment of the explosion, the remnant formed is a NS. In this example, after the SN the semi-major axis of the orbit increases (to $\sim 2500~R_\odot$), as well as the orbital eccentricity ($e \sim 0.7$). The secondary in turn evolves after a short time ($2~\mathrm{Myr}$) given the mass gained during the first episode of mass transfer. When helium fusion in the core ignites, the secondary enters the giant phase and CE starts $\sim 2~\mathrm{Myr}$ later, due to the prior eccentricity of the orbit. At the end of the CE phase, the secondary is stripped of its envelope and the semi-major axis is reduced to $\sim 2.5~R_\odot$, while friction during the CE circularizes the orbit. Because of the short orbital separation, a continuous \textit{case BB} mass transfer from the stripped star onto the NS occurs \citep{2015MNRAS.451.2123T}. Shortly after, the secondary explodes in an ultra-stripped SN and forms the second NS of the system. Because of the reduced NS kick, the system is given a new initial semi-major axis ($\sim 4~R_\odot$) and eccentricity ($e \simeq 0.4$) but has increased chances of survival. This system eventually merges due to the emission of GWs after $\sim 5~\mathrm{Gyr}$. We observe that for systems that follow this track, the stellar evolution time is negligible compared to the typical time for merger due to emission of GWs (see Figs.~\ref{fig:bns_properties_3_mets}~and~\ref{fig:bns_properties_3_channels}).

To summarize, this evolutionary track  features a phase of CE between a giant and a NS, and a later phase of mass transfer between the stripped star and the NS (\textit{case BB} mass transfer). As will be discussed in Sect.~\ref{sec:discussion}, this track does not produce as many systems as the other evolutionary tracks with our choice of model parameters, in particular at high metallicities.

\subsubsection{Equal-mass ratios: Co-evolution}
\label{sec:equal_mass_evol_track}

\begin{figure}
    \centering
    \resizebox{\hsize}{!}{\includegraphics{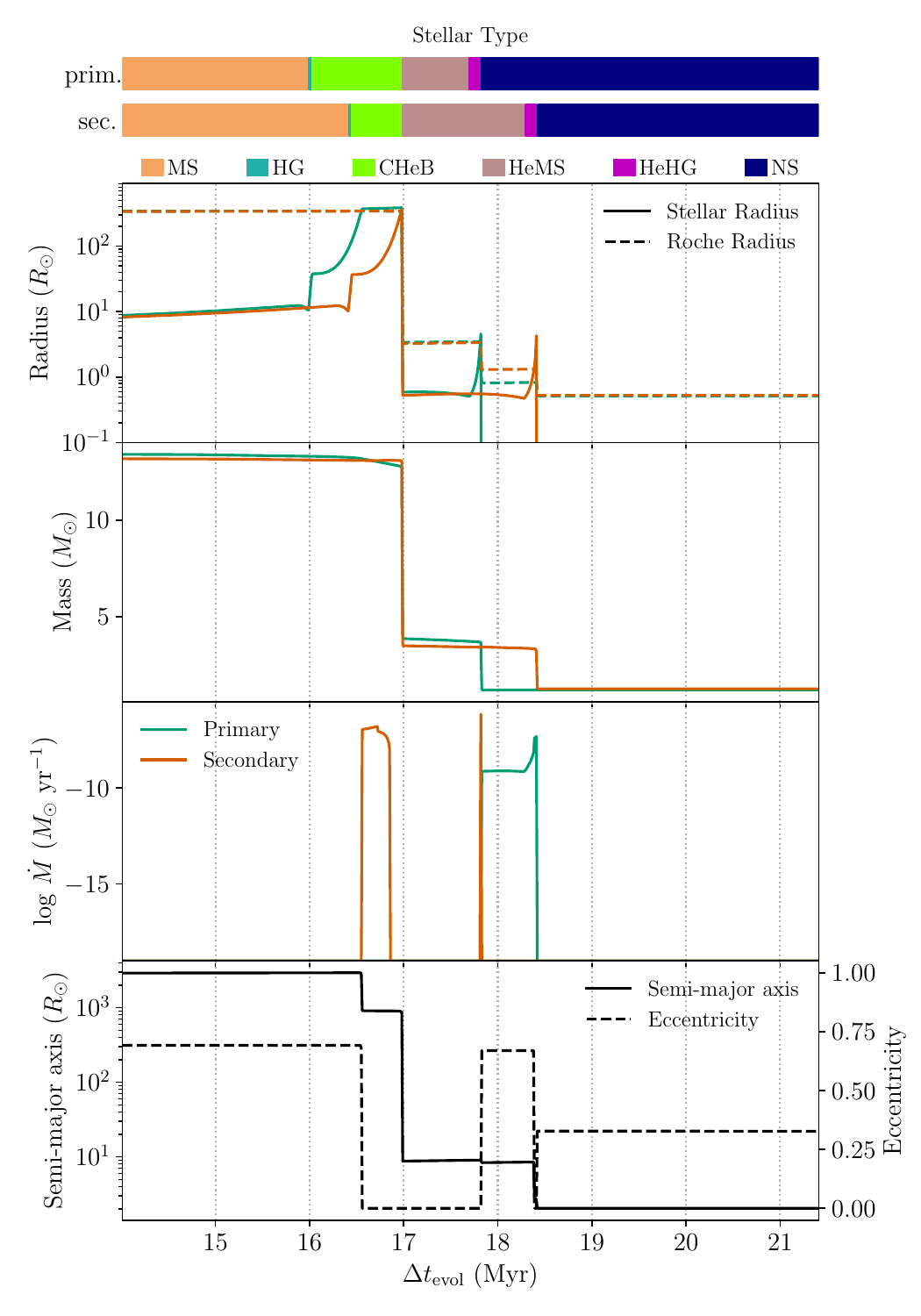}}
    \caption{Evolution of a binary system on the \emph{equal-mass} evolutionary track forming a BNS at $Z = 0.0065$. The first $14~\mathrm{Myr}$ of main sequence evolution are hidden. See Fig.~\ref{fig:standard_evol_example} for panel description.}
    \label{fig:equal_mass_evol_example}
\end{figure}

At all metallicities, many BNS systems are produced from the evolution of binary systems whose ZAMS star masses were almost equal ($q \simeq 1$). In the example shown in Fig.~\ref{fig:equal_mass_evol_example}, $M_1 = 13.44~M_\odot$ and $M_2 = 13.21~M_\odot$, i.e. $q = 0.98$. The semi-major axis is initially of $\sim 2900~R_\odot$ and the eccentricity $e \simeq 0.69$. Both stars evolve on similar timescales and both are in their CHeB phase after $16.44~\mathrm{Myr}$. Due to the expansion of both stellar envelopes, and to the slightly higher mass, and thus radius of the primary, a phase of stable mass transfer is initiated, leading to the circularization of the orbit and therefore of the reduction of the semi-major axis. Both stars enter in contact and a joint phase of CE occurs, where the two helium cores orbit within the common envelope. After the envelope ejection, the helium cores have a separation of $\sim 8.7~R_\odot$, and most of the mass has been lost by the binary. The primary is slightly more massive than the secondary, therefore evolves first, loses some of its mass to the secondary by \textit{case BB} mass transfer, and explodes in an ultra-stripped SN. Then the secondary evolves, stable mass transfer onto the NS follows (\textit{case BB} mass transfer), until the secondary explodes in another ultra-stripped SN. 

Several variations of the example described above and shown in Fig.~\ref{fig:equal_mass_evol_example} can be found within this evolutionary track, and correspond to $9$ evolutionary \emph{sequences} (see Sect.~\ref{sec:tracks}), but they all share the same distinctive feature of this evolutionary track: a quasi-equal initial mass ratio, leading to a synchronized evolution of both stars and a phase of CE with the two stars on their giant phase. Variations depend on the exact stellar types at the moment of the CE (CHeB or EAGB) and the stellar types of the remaining helium cores at the end of CE. In the cases where the primary is on the EAGB at the onset of CE, the remaining helium cores after CE are already on the HG. A second source of variation is the nature of the stellar type of the helium stars at the moments of the SN. Final variations are found depending on the metallicities, but again, share the same distinct features of this evolutionary track. Note that binary systems in this track typically evolve on a faster timescale than the others, due to the joint evolution of the primary and the secondary. The time between binary formation at ZAMS and BNS formation ranges between $\sim 10$ and $100~\mathrm{Myr}$.

This evolutionary track is also discussed in the literature, for example in \citet{2018MNRAS.481.4009V} where it is labelled Channel II; \citet{2019MNRAS.490.3740N} where it is labelled Channel III and \citet{2021MNRAS.508.5028B, 2023MNRAS.524..426I} where it is labelled Channel IV. This evolutionary track was also discussed in various contexts by e.g. \citet{1995ApJ...440..270B, 1998ApJ...506..780B, 2006MNRAS.368.1742D, 2015ApJ...806..135H}. Note that the existence of this evolutionary track is more subject to our choice of model parameters than the others. Indeed, because we impose $q_\mathrm{crit} = +\infty$ for mass transfers from stripped stars, this evolutionary track only features one phase of CE. With other choices of parameters, there would be additional variations with one or two additional phases of CE with the helium cores, increasing the amount of systems with mergers before the BNS formation and therefore reducing the efficiency of this evolutionary track.

\subsubsection{Accretion-induced collapse}
\label{sec:aic_evol_track}

\begin{figure}
    \centering
    \resizebox{\hsize}{!}{\includegraphics{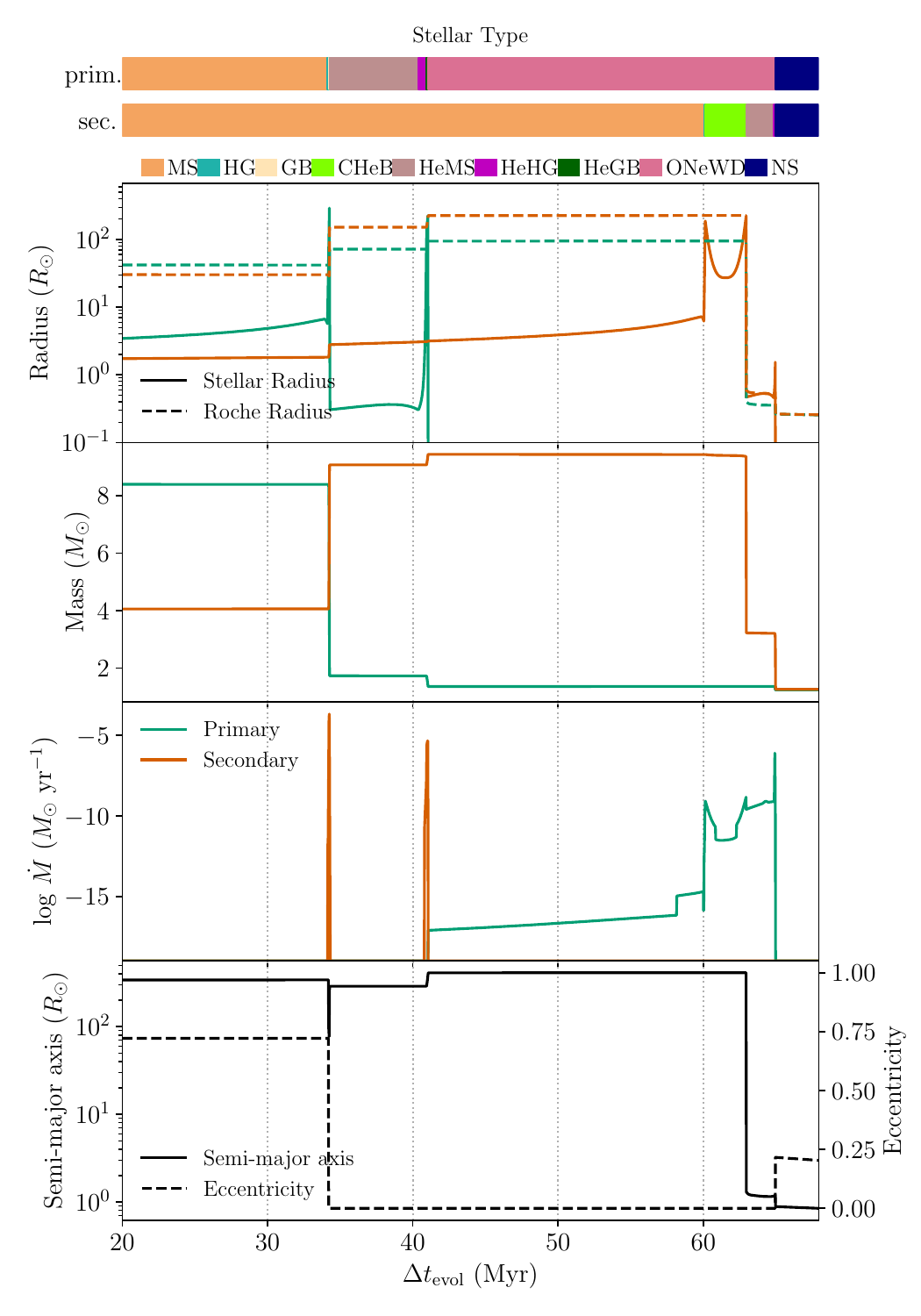}}
    \caption{Evolution of a binary system on the \emph{AIC} evolutionary track forming a BNS at $Z = 0.00044$. The first $20~\mathrm{Myr}$ of main sequence evolution are hidden. See Fig.~\ref{fig:standard_evol_example} for panel description.}
    \label{fig:aic_example}
\end{figure}

The final evolutionary track found with this model features the AIC of a WD into a NS. This evolutionary track has been discussed in \citet{2020RAA....20..135W}, where the authors show in particular the limited parameter space at ZAMS ($P_\mathrm{orb}$, $M_1$, $M_2$) within which this process can occur. \citet{2018MNRAS.474.2937C} show that this evolutionary track could contribute to the BNS merger rate, owing to lower natal kicks, while \citet{2012ApJ...759...52D} estimate its contribution to $8\%$ of the total BNS formed at solar metallicity. AIC of WDs is also discussed in a more general context in e.g. \citet{2009MNRAS.395.2103M, 2010A&A...515A..88W}. In our case, this evolutionary track contains two evolutionary sequences as defined in Sect.~\ref{sec:tracks}. One of them is dominant at the lowest redshifts and differs from the other one only by one element: the primary does not enter the FGB. The physical properties of both evolutionary sequences are however similar. We show an example of a representative system in Fig.~\ref{fig:aic_example}. In this evolutionary track, the initial mass ratio can reach much lower values than in the other tracks, $0.5 \lesssim q < 1$. In this example, $q \simeq 0.48$. The initial mass of the secondary is lower than the typical mass of NS progenitors (here $M_2 = 4.06~M_\odot$), due to the first phase of mass transfer.

The primary evolves first from the MS after $34~\mathrm{Myr}$. When it reaches the HG, a phase of stable mass transfer starts, reducing the semi-major axis and circularizing the orbit. During this event of mass transfer, the primary evolves and loses more mass to the companion. Because the mass transfer is not entirely conservative (some mass is lost by the binary), the semi-major axis increases again, until the primary has lost its entire hydrogen envelope and becomes a stripped helium star. All these events occur on time scales $\lesssim 0.1~\mathrm{Myr}$. At that point, the secondary has a mass that is higher than the initial primary mass. The primary helium star evolves (here in $\sim 6~\mathrm{Myr}$), and a new phase of mass loss can occur. When the primary enters the NHeGB, the remaining core mass is therefore much lower than in the other scenarios (here $1.36~M_\odot$). At the moment of the SN, the primary collapses into an oxygen-neon WD. As all the mass of the helium core forms the WD, no natal kick is imparted to the WD. The orbit thus remains circular. In turn, the (now massive) secondary leaves the MS and evolves, roughly $20~\mathrm{Myr}$ later. When it reaches the HG, a phase of CE is triggered, with the WD. After the envelope ejection, the orbital separation is reduced to an extremely low value (here $1.3~R_\odot$). Throughout the evolution of the helium star, stable mass transfer occurs and gradually increases the mass of the WD. When it reaches the Chandrasekhar mass  by accretion, the WD collapses and forms a NS of $1.24~M_\odot$ (this value is unique for all systems by construction in COSMIC but would nevertheless be expected to have a low dispersion). Just after, the secondary in turn collapses into a NS thanks to the mass gained during the first phase of mass transfer. The BNS system formed that way has an extremely small initial semi-major axis and merges in typically a few tens of $\mathrm{Myr}$. 

\vspace{0.5cm}
In the context of BBH and NSBH formation history, several authors discuss an additional evolutionary track where no phase of CE is triggered, and all mass transfers are dynamically stable \citep{2019MNRAS.490.3740N, 2021MNRAS.508.5028B, 2023MNRAS.524..426I}. In this study, we put our focus on BNS systems and do not find that such an evolutionary track contributes significantly to the population of merging BNSs. This is expected \citep[see e.g.][]{2023MNRAS.524..426I}, since in the case of BBH and NSBH progenitors, their higher masses allow for longer phases of mass transfer, which can shrink the orbit significantly. Additionally, because of the higher remnant masses, systems with higher initial semi-major axes have a lower probability to be disrupted by either of the SNe: this allows detached binaries which evolve without a phase of CE to still produce compact object remnants that remain gravitationally bound.

\subsection{Properties of stellar progenitors}\label{sec:fixed_met_progenitors}

\begin{figure*}
    \centering
    \includegraphics[width=\textwidth]{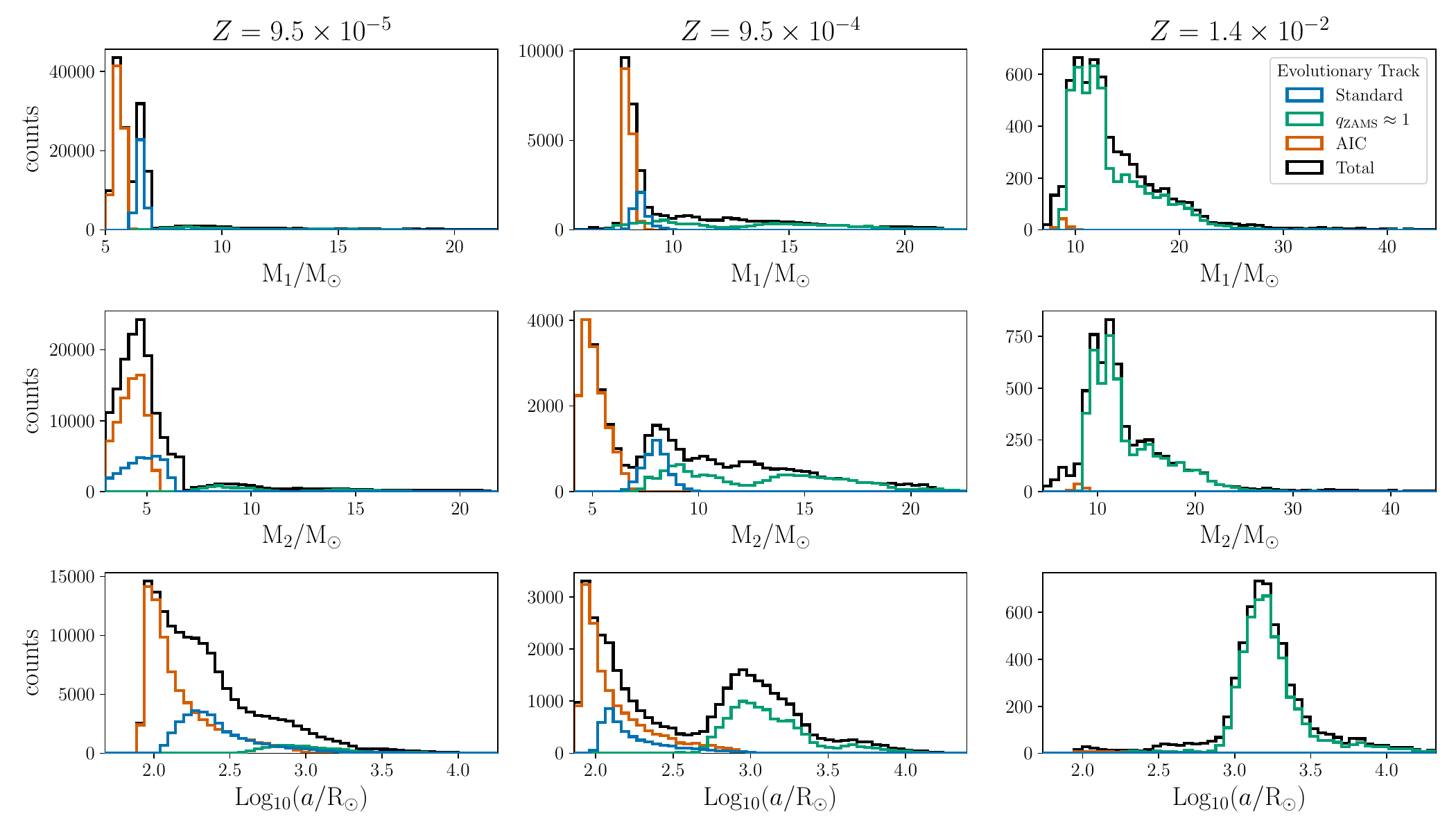}
    \caption{Distribution of the properties of the population of BNS progenitors, at their formation (ZAMS) and at three metallicities, $Z = 9.5 \times 10^{-5}$ (left); $Z = 9.5 \times 10^{-4}$ (center); and $Z = 1.4 \times 10^{-2}$ (right). From top to bottom, the three rows show the distributions of (1) primary masses; (2) secondary masses; (3) initial semi-major axes. The \emph{standard} evolutionary track is shown in blue; the \emph{equal-mass} track in green; and the \emph{AIC} track in orange. In black, we show the total population.} 
    \label{fig:initial_properties_3_mets}
\end{figure*}

We now focus on the properties of the stellar progenitors at fixed metallicity. As could be expected from the discussion in Sect.~\ref{sec:results_evol_tracks}, they differ across the three evolutionary tracks. In Fig.~\ref{fig:initial_properties_3_mets}, we show the distributions of initial binary properties at ZAMS for the metallicities $Z = 9.5 \times 10^{-5}$, $Z = 9.5 \times 10^{-4}$ and $Z = 1.4 \times 10^{-2}$, for each evolutionary track. 

In this figure, we filter the simulated populations at each metallicity such that $\Delta t_\mathrm{delay} < t_\mathrm{Hubble}$ to simplify our comparative studies and to only focus on systems which may merge within the Hubble time. Because binaries with higher metallicities are more likely to form at lower redshift (see Eq.~(\ref{eq:metalmean})), the tails of these populations with the longest merger times do not contribute to the population of BNSs which merged through cosmic time until today, even if their merger time is lower than the Hubble time. The merger rates (Fig.~\ref{fig:merger_rate}) are computed using the exact shapes of these distributions. By construction (see Eq.~(\ref{eq:merger_rate})), they do not include the systems such that $t(z_\mathrm{ZAMS}) + \Delta t_\mathrm{delay} > t_\mathrm{Hubble}$, where $t(z_\mathrm{ZAMS})$ is the time at which the progenitor binary is formed.

While the distribution of the various binary properties at ZAMS vary across metallicities, several features are common. We observe in the first and second columns of Fig.~\ref{fig:initial_properties_3_mets} that the \emph{AIC} evolutionary track takes place for quite narrow mass ranges of the primary due to the restrictive conditions for AIC to occur, and slightly wider ranges for the secondary. As expected, these masses are on the lower end, around $5~M_{\odot}$, since these stars are not massive enough to collapse directly into a NS. The \emph{equal-mass} track occurs primarily for higher-mass progenitors, with the mass distribution extending beyond $20~M_{\odot}$, while the mass distributions for the \emph{standard} track falls between the \emph{equal-mass} track and \emph{AIC}. We also notice that for all tracks the mass distribution shifts to higher values at higher metallicity. This effect is to the first order due to enhanced stellar winds at those metallicities: to form NS progenitors of similar masses right before core-collapse, a higher-metallicity star must have a larger mass at ZAMS than a lower-metallicity star. In principle, overshooting of the convective envelope into the core may also play a role here, but this is modelled in COSMIC as in \citet{1998MNRAS.298..525P}, without mass or metallicity dependence.

\begin{figure}
    \centering
    \resizebox{\hsize}{!}{\includegraphics{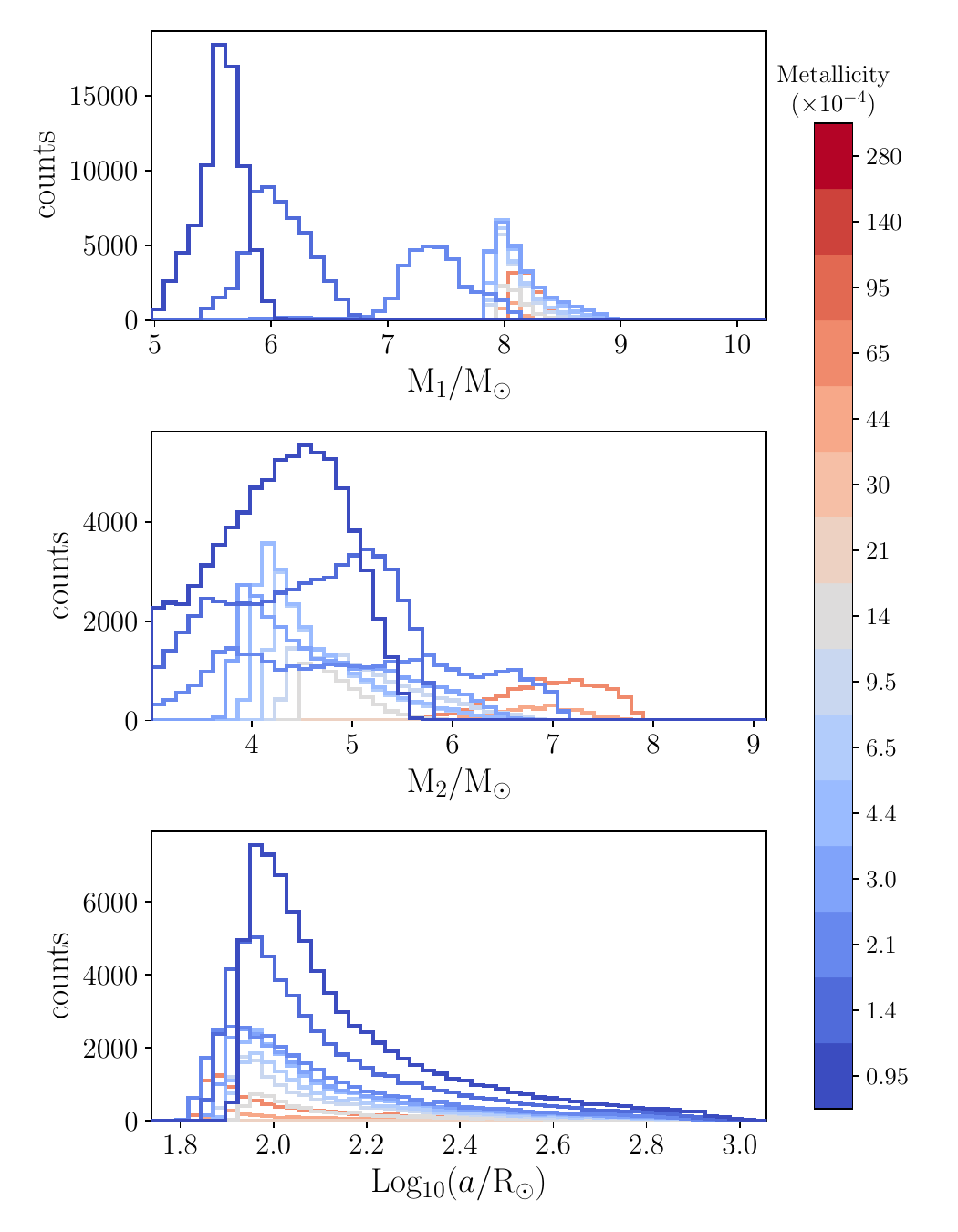}}
    \caption{Distributions of stellar progenitors properties at ZAMS for the \emph{AIC} evolutionary track and for various metallicities. From top to bottom, the three panels show the distributions of (1) primary masses; (2) secondary masses; (3) initial semi-major axes.}
    \label{fig:initial_aic}
\end{figure}

The \emph{AIC} track disappears almost completely at solar metallicity, as can be seen on the right column of Fig.~\ref{fig:initial_properties_3_mets}. In fact, as we show in Fig.~\ref{fig:merger_rate} it becomes subdominant to the other tracks at $z\lesssim 3.5$. On the other hand, the \emph{equal-mass} evolutionary track dominates at higher metallicities (see the right column of Fig.~\ref{fig:initial_properties_3_mets}). To further explore this effect for the \emph{AIC} track, we show the evolution of the progenitor properties at ZAMS  with metallicity in Fig.~\ref{fig:initial_aic}. In fact, at a given metallicity, if $M_1$ is too high, the primary directly collapses into a NS and if $M_1$ is too low, it collapses into a carbon-oxygen WD which cannot undergo AIC. The range of masses of oxygen-neon WD progenitors (necessary for AIC events) is therefore quite narrow. When metallicity increases, the mass lost by the primary before its core collapses also increases, thus requiring higher initial ZAMS masses $M_1$ to produce oxygen-neon WDs. Indeed, as we see in Fig.~\ref{fig:initial_aic}, the primary mass increases with metallicity until it stalls around $8~M_\odot$. The reason this trend stops is due to the transition between the two evolutionary sequences that compose the \emph{AIC} evolutionary track (see Sect.~\ref{sec:aic_evol_track}). At these higher metallicities, it is this time the secondary whose mass $M_2$ gradually increases with metallicity. The reason is again related to the mass losses from stellar winds: the secondary initial mass must be higher at higher metallicities. The number of systems produced in the \emph{AIC} evolutionary track gradually decreases when metallicity increases. This is partially explained by the need for increasing progenitor masses at higher metallicities, which therefore lie further on the decreasing IMF \citep{2001MNRAS.322..231K}. Combined with the very restrictive mass range of the progenitors of oxygen-neon WDs, and other potential effects linked to binary evolution, this progressively reduces the contribution of the \emph{AIC} evolutionary track to the total as metallicity increases.

The distribution of progenitor properties at ZAMS for the \emph{standard} and the \emph{equal-mass} evolutionary tracks are shown in Appendix~\ref{ap:initial_properties}. The \emph{standard} evolutionary track has two components, with similar ZAMS semi-major axes. At lower metallicities, the initial masses $M_1$ and $M_2$ are lower and reach up to $M_1 \lesssim 9~M_\odot$; $M_2 \lesssim 8~M_\odot$ at higher metallicities. Above $Z \simeq 0.1~Z_\odot$, this track almost vanishes. The distribution of progenitor masses for the \emph{equal-mass} track does not exhibit any particular trend. 

Another interesting feature concerning the BNS progenitors is the range of initial semi-major axes. In the first and second columns of Fig.~\ref{fig:initial_properties_3_mets}, we observe that the \emph{AIC} track occurs for the lowest progenitor semi-major axes, while the \emph{equal-mass} track takes places for larger semi-major axes of $\sim 1000\,R_{\odot}$. These findings are in agreement with the progenitor ranges discussed for the \emph{AIC} track by \citet{2020RAA....20..135W}, namely $M_1 \sim 6 - 10~M_\odot$; $q \sim 0.2 - 0.4$ and $P \sim 400 - 1000~\mathrm{days}$. We also see the impact of metallicity as discussed previously. 

Finally, we also note that the initial eccentricity does not affect the tracks followed by the system: the distribution of eccentricities for the progenitors which successfully produce BNSs that merge within the Hubble time is the same as the distribution assumed for the entire population. A possible application of these results would be to find general criteria that would determine which evolutionary track a given binary is more likely to follow \citep[see also the discussion in][]{2019MNRAS.490.5228B}. This study is however beyond the scope of the present work and we leave it for future investigations.

\subsection{Properties of binary neutron star populations}\label{sec:bns-pops}

\begin{figure*}
    \centering
    \includegraphics[width=\textwidth]{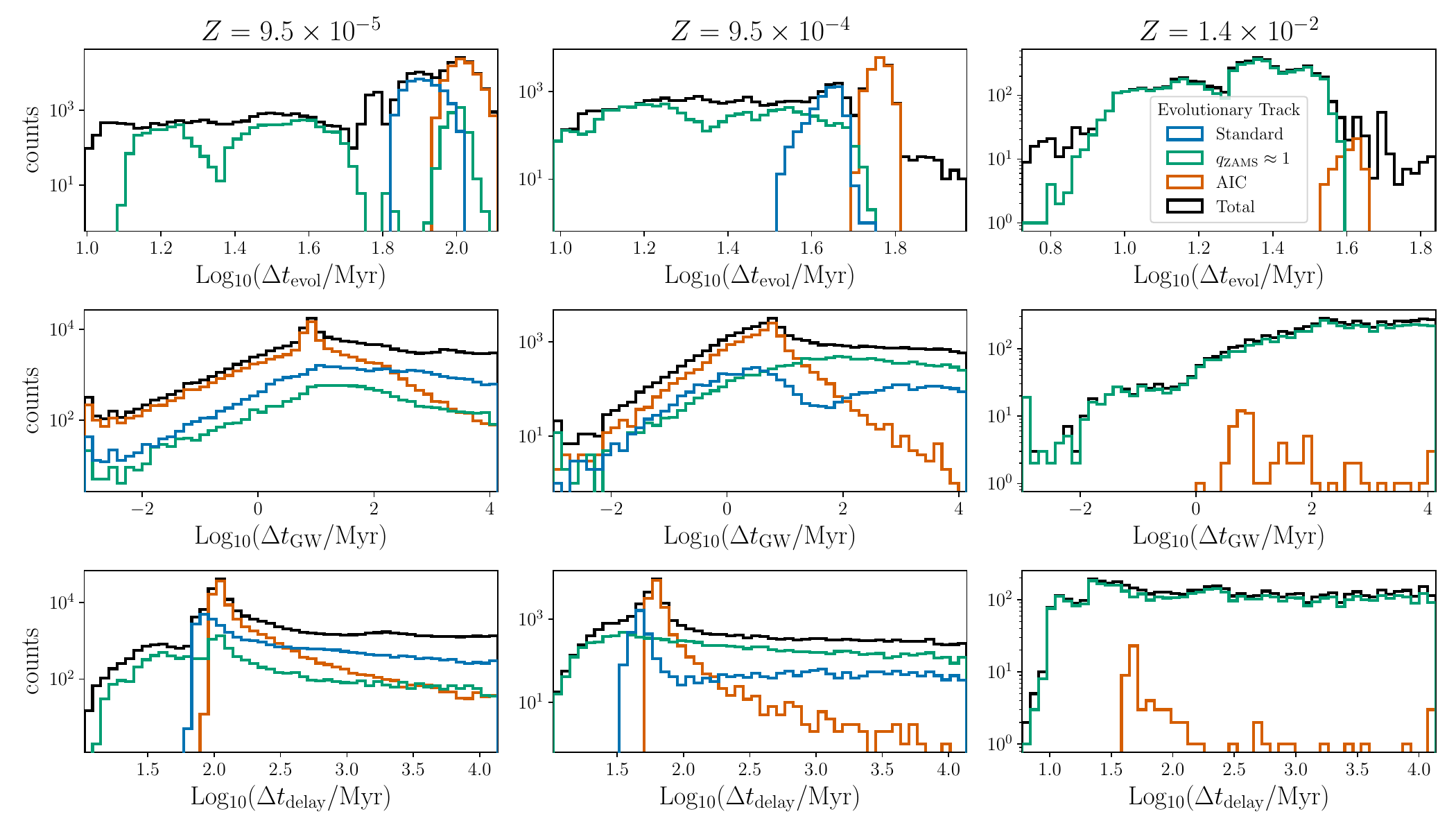}
    \caption{Distribution of the properties of the population of BNS, at their formation, i.e. right after the secondary SN, and at three metallicities, $Z = 9.5 \times 10^{-5}$ (left); $Z = 9.5 \times 10^{-4}$ (center); and $Z = 1.4 \times 10^{-2}$ (right). From top to bottom, the three rows show the distributions of (1) stellar evolution times; (2) merger times; (3) delay times. As in Fig.~\ref{fig:initial_properties_3_mets}, the \emph{standard} evolutionary track is shown in blue; the \emph{equal-mass} track in green; and the \emph{AIC} track in orange. In black, we show the total population. The count sum of the three evolutionary tracks accounts for $\sim 95\%$ of systems (see Sect.~\ref{sec:tracks}).} 
    \label{fig:bns_properties_3_mets}
\end{figure*}

We now discuss the properties of BNS systems at their formation, i.e. right after the secondary SN, for the three evolutionary tracks. We focus on three parameters: Stellar evolution timescale $\Delta t_\mathrm{evol}$ (time from the formation of the stellar binary (ZAMS) to the formation of the BNS), GW orbital decay timescale $\Delta t_\mathrm{GW}$ (time from BNS formation to merger) and the total delay time $\Delta t_\mathrm{delay} = \Delta t_\mathrm{evol} + \Delta t_\mathrm{GW}$. These parameters are shown in Fig.~\ref{fig:bns_properties_3_mets}.

From Fig.~\ref{fig:bns_properties_3_mets} it is immediately clear that, for individual metallicities, the distribution of delay times does not follow the power-law scaling of $1 / t$ as is often assumed in the literature (e.g. \citealt{2018MNRAS.474.2937C,2019MNRAS.490.3740N}). In fact, depending on the metallicity and on which evolutionary track is dominant, the distribution of delay times can be flat (see the right column of Fig.~\ref{fig:bns_properties_3_mets}) or extremely peaked (see the left column of Fig.~\ref{fig:bns_properties_3_mets}, where the counts are shown in logarithmic scale). The \emph{equal-mass} evolutionary track even shows multiple peaks, which are due to the $9$ different evolutionary \emph{sequences} that constitute this track (see Sect.~\ref{sec:tracks}), and which appear for specific values of $M_1$ at ZAMS. Moreover, a significant fraction of binaries has a delay time dominated by the stellar evolution timescale, with extremely short merger times. This is the case notably for the \emph{AIC} evolutionary track, for which the delay times due to GW emission are extremely short, of the order of $10$~Myr, and the total delay time is therefore dominated by the stellar evolution timescale, of the order of $100$~Myr. This effect occurs because the distribution of BNS semi-major axes at formation for the \emph{AIC} track is very peaked around low values, below $R_{\odot}$. On the contrary, systems formed via the \emph{equal-mass} evolutionary track are born with much larger semi-major axes and their delay time is therefore dominated by the orbital decay due to emission of GWs.

Another observation is that the different evolutionary tracks are well distinguished by the progenitor binary evolution time $\Delta t_\mathrm{evol}$. Progenitors with equal-mass ratios evolve more rapidly due to their initially higher masses (see the green distributions in Fig.~\ref{fig:initial_properties_3_mets}, first two rows and Fig.~\ref{fig:bns_properties_3_mets}, first row), BNSs formed through the \emph{standard} evolutionary track have slightly longer stellar evolution times, while the longest evolution times are required for the \emph{AIC} evolutionary track, mostly due to the smaller total mass in the system.

The distributions of delay times $\Delta t_\mathrm{delay}$ all have a minimum at $\sim 1~\mathrm{Myr}$ corresponding to systems with the shortest evolution times and rapid mergers ($< 10^{-1}~\mathrm{Myr}$). Conversely, the maximum delay time is fixed to be $t_\mathrm{Hubble}$, though all these distributions extend to longer delays. The peaks of the distributions for each evolutionary track are primarily affected by the intensities and orientations of the NS natal kicks, but also by events of stellar evolution, such as CE efficiency, \textit{case BB} mass transfer rates, or the radius expansion of stripped stars. A careful study of the impact of these parameters on the delay time distributions properties would require a systematic variation of those parameters and is outside of the scope of this paper.

\begin{figure}[!b]
    \centering
    \resizebox{\hsize}{!}{\includegraphics{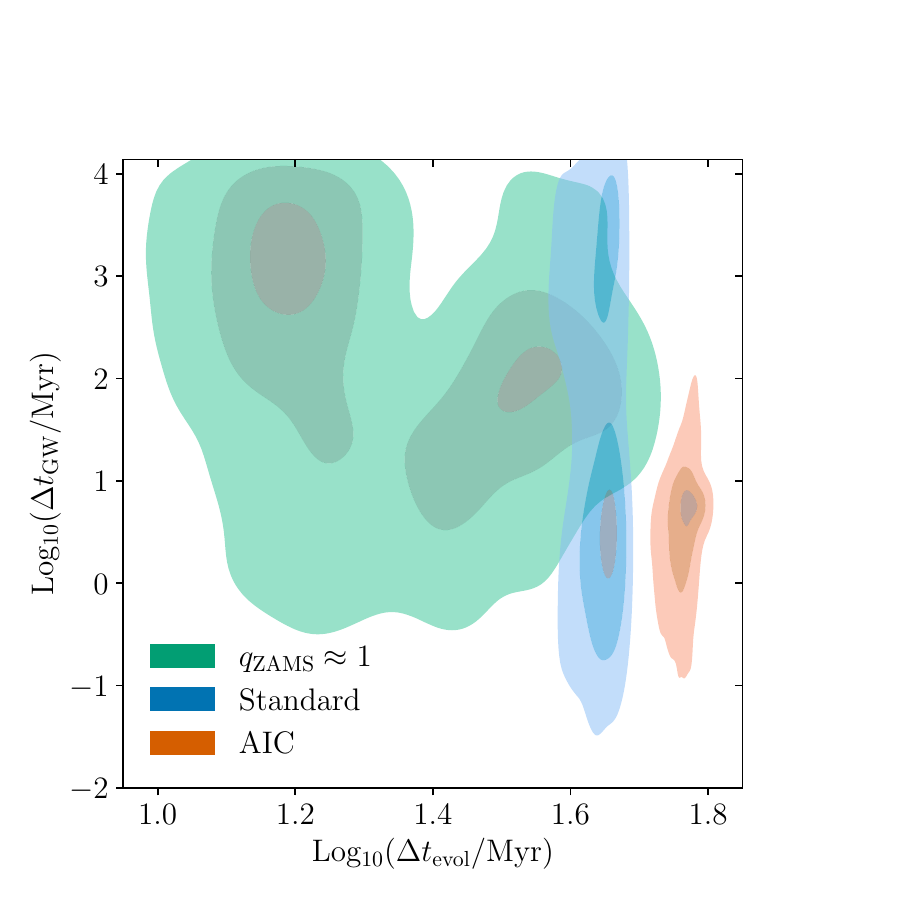}}
    \caption{Correlation between the formation and merger times for the three evolutionary tracks, at $Z = 9.5\times 10^{-4}$, and for systems with $\Delta t_\mathrm{delay} < t_\mathrm{Hubble}$. The three tracks are found on distinct regions of the plane, as discussed in Sect.~\ref{sec:bns-pops}. The shaded contours represent the smallest regions containing $10\%$, $50\%$ and $90\%$ of the systems.} \label{fig:00095_corner_plot}
\end{figure}

To complement the information in Fig.~\ref{fig:bns_properties_3_mets}, we show the joint distributions of the stellar evolution timescale $\Delta t_\mathrm{evol}$ and the GW orbital decay timescale $\Delta t_\mathrm{GW}$ at a given metallicity $Z = 9.5 \times 10^{-4}$ in Fig.~\ref{fig:00095_corner_plot}. This joint representation allows to better visualize the correlation between these two quantities for the three evolutionary tracks. The \emph{standard} evolutionary track has long evolution times, but not the longest, and a distribution of merger times that ranges between $< 1~\mathrm{Myr}$ and $t_\mathrm{Hubble}$. The \emph{equal-mass} evolutionary track has short evolution times and long merger times, while the \emph{AIC} evolutionary track has the longest evolution times and short merger times.

These results could have profound effects on the properties of BNS populations across cosmic times. Since dominant tracks vary as a function of metallicity, we may expect the BNS properties (in particular their delay time) to also vary with metallicity, and hence redshift. In other words, the merging BNS population observed in the local Universe may not be representative of the higher redshift population that merged at an earlier epoch. We further discuss this idea in Sect.~\ref{sec:discussion}. 

\begin{figure*}
    \centering
    \includegraphics[width=\textwidth]{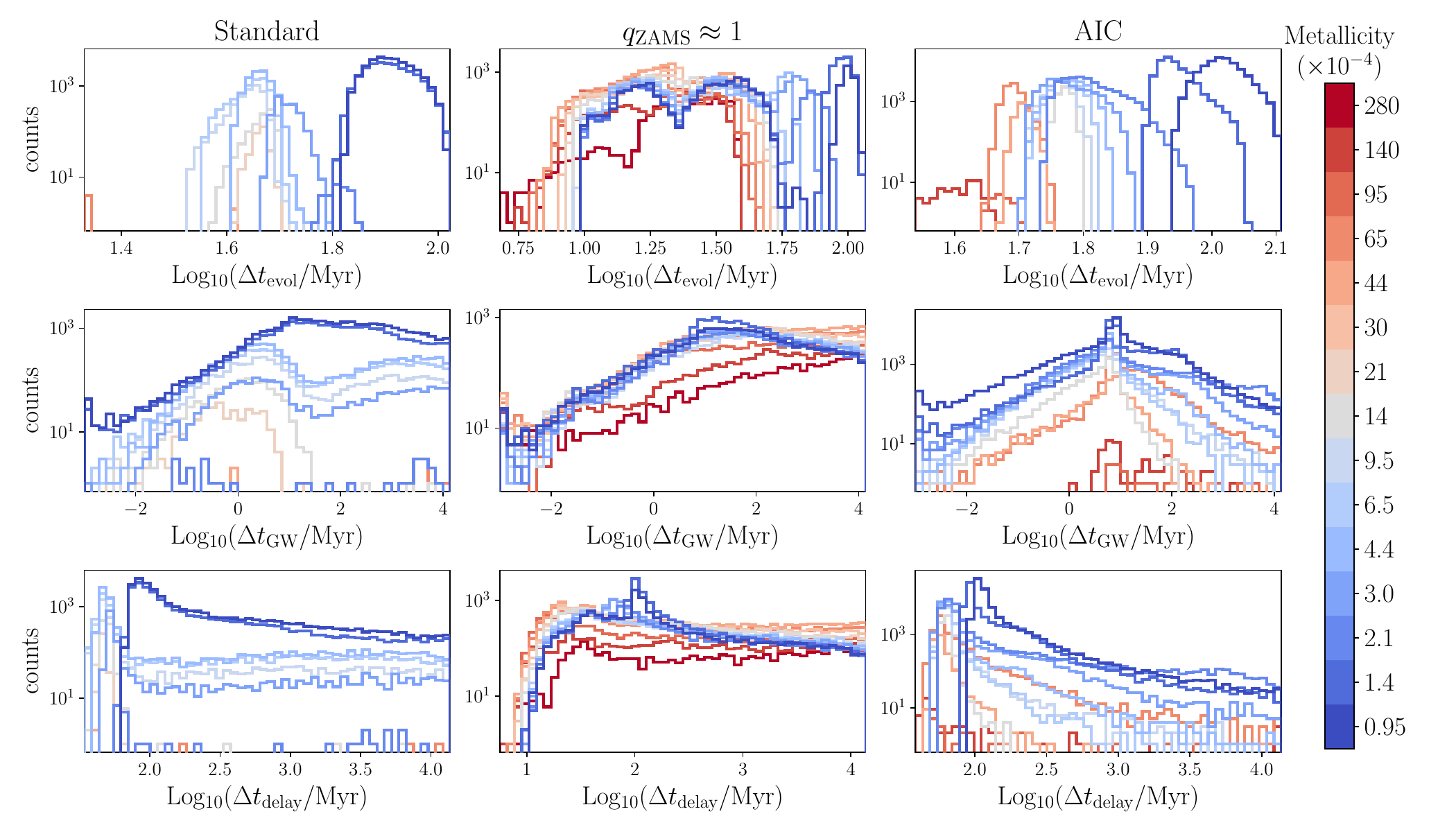}
    \caption{Distribution of the properties of the population of BNS, at their formation, i.e. right after the secondary SN for the \emph{standard} evolutionary track (left), the \emph{equal-mass} track (center), and the \emph{AIC} track (right). Each color represents the population at one of the sampled metallicity values. From top to bottom, the three rows show the distributions of (1) stellar evolution times; (2) merger times; (3) delay times.}
    \label{fig:bns_properties_3_channels}
\end{figure*}

The evolution of BNS properties as a function of progenitor metallicity is presented in Fig.~\ref{fig:bns_properties_3_channels}. A general trend that we observe is the decrease in the efficiency of BNS formation as metallicity increases. This is mostly due to the increased stellar wind intensities which push progenitors to higher ZAMS masses, further in the tail of the initial mass function distribution. Overall, about one system in $10^4 - 10^6$ produces a BNS which merges within the age of the Universe.

On the left column in Fig.~\ref{fig:bns_properties_3_channels}, we observe again that the \emph{standard} evolutionary track is far less efficient at high metallicities. We confirmed that the \emph{standard} track produces systems with $\Delta t_\mathrm{delay} > t_\mathrm{Hubble}$, all the more at higher metallicities. These systems are removed in this study. We also observe that the distribution of initial semi-major axes of the BNSs does not evolve with metallicity, however the stellar evolution timescale decreases with metallicity. This is expected, since the initial stellar masses of BNS progenitors are higher at higher metallicity, which leads to quicker stellar evolution timescales. We also stress that even in the \emph{standard} evolutionary track, the distribution of time delays does not follow the $1/t$ scaling. Indeed, the distribution of initial BNS semi-major axes is not log-normal, and moreover, the stellar evolution timescale is of the same order or longer than the GW timescale, for a significant fraction of the systems.

We now examine the \emph{equal-mass} track, shown on the central column of Fig.~\ref{fig:bns_properties_3_channels}. In this case the distributions of timescales are much more complex, in particular the distribution of stellar evolution times. As we mentioned in Sect.~\ref{sec:equal_mass_evol_track}, this track is composed of $9$ evolutionary \emph{sequences}, which in part leads to the observed complexity, because their relative contributions to the total varies with metallicity. This evolutionary track produces BNS systems efficiently at all metallicities and therefore dominates among the (local) BNS events observed with GW detectors, since the \emph{standard} and \emph{AIC} tracks are inefficient at higher metallicities.

The \emph{AIC} evolutionary track is shown on the right column of Fig.~\ref{fig:bns_properties_3_channels}. One major difference between this track and the other two is the distribution of merger times. In the \emph{AIC} track, the semi-major axis after the SN of the secondary is always extremely reduced after the phase of CE with the WD. After the final phases of mass transfer from the helium companion onto the WD that lead to its AIC, and the ultra-stripped SN of the secondary, the semi-major axis still remains extremely small in most cases, leading to a merger in typically $\lesssim 10~\mathrm{Myr}$. This property does not evolve with redshift. We note that the radius evolution of stripped stars is uncertain at low metallicities, and we thus expect these results to be dependent on our choice of model parameters. Only the overall number of BNSs formed by this evolutionary track decreases at higher metallicities. The \emph{standard} evolutionary track has much longer merger times at low metallicities, while for the \emph{equal-mass} evolutionary track, the merger time tends to increase with metallicity.

\subsection{Merger rate density}\label{sec:merger-rate}

\begin{figure}
    \centering
    \resizebox{\hsize}{!}{\includegraphics{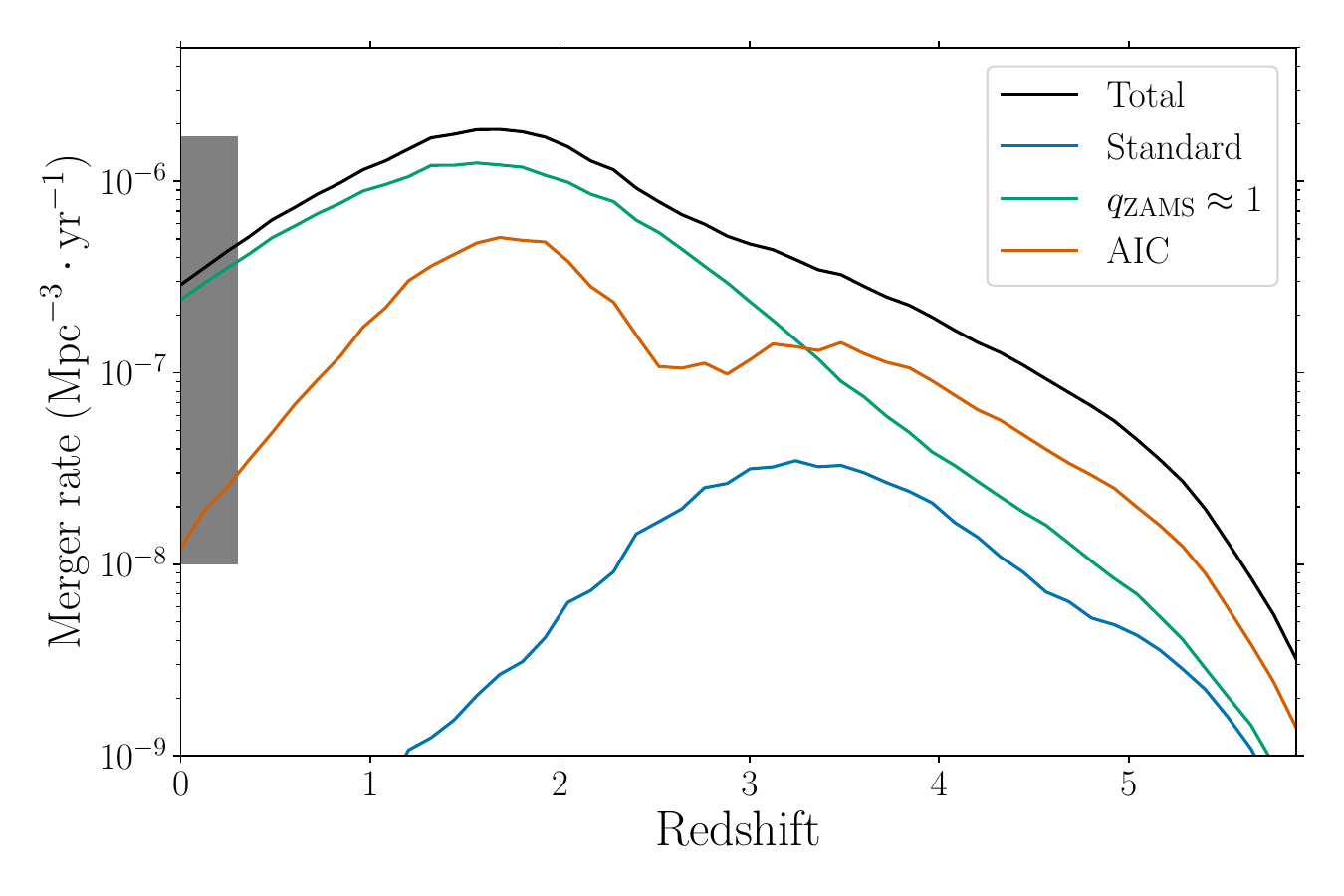}}
    \caption{Evolution of the merger rate with redshift: total rate (black); \emph{standard} evolutionary track (blue; see Sect.~\ref{sec:standard_evol_track}); \emph{equal-mass} progenitors (green; Sect.~\ref{sec:equal_mass_evol_track}); and \emph{AIC} (orange; see Sect.~\ref{sec:aic_evol_track}). The relative contribution of each track to the merging population varies with redshift, with the \emph{AIC} track prominent at high redshifts, and the \emph{equal-mass} track dominant at low redshifts. The grey box represents the constraint on the merger rate in the local Universe after the O3 observing run of LIGO/Virgo \citep{2023PhRvX..13a1048A}. It is stretched up to $z = 0.3$ for visualisation purposes.}
    \label{fig:merger_rate}
\end{figure}

We present in Fig.~\ref{fig:merger_rate} the merger rate obtained after combining all the metallicity bins following the procedure described in Sect.~\ref{sec:merger_rate_density}, along with the constraint on the local merger rate from GW observations: $\mathcal{R}_\mathrm{merg}(z=0) = 10 - 1700 ~\mathrm{Gpc^{-3}\cdot yr^{-1}}$ \citep{2023PhRvX..13a1048A}. Our model is compatible with the observed local merger rate constraint, although we note that it still covers a broad range. In that regard, new detections will help constraining population models.

We now focus on the contributions from each of the three evolutionary tracks. First, we observe that the \emph{standard} evolutionary track contributes to a very small fraction of the merger rate at all redshifts. Moreover, the evolutionary track dominating the population of merging binaries changes with redshift: at higher redshifts, most mergers are produced by BNS systems formed via the \emph{AIC} evolutionary track, while at lower redshifts, the population we see merging in the local Universe is mostly formed by equal-mass ZAMS progenitors. This result suggests that the BNS population observable with GWs may not be representative of the entire merging BNS population across cosmic times.

The two peaks in the merger rate contribution from the \emph{AIC} evolutionary track originate from the two evolutionary sequences that contribute to this track that we discussed in Sect.~\ref{sec:aic_evol_track}: one at low metallicities and the other at higher metallicities. In the transition between the two peaks, the evolutionary sequence where the primary goes through the FGB (higher redshifts) is not as efficient, and the second sequence where the primary does not enter the FGB phase is not yet efficient. 

Because most merging systems at high redshift are formed by the \emph{AIC} evolutionary track, they have very short merger times $\Delta t_\mathrm{GW}$. The merger rate density shown in Fig.~\ref{fig:merger_rate} therefore closely follows the evolution of the SFR density. 

Finally, it is interesting to note that the total merger rate evolution is quite smooth and hides the diversity of progenitor evolutionary tracks that make up the BNS population, and despite the broad diversity of the distribution shapes shown in Fig.~\ref{fig:bns_properties_3_mets}.

\section{Discussion}\label{sec:discussion}

We explored the formation scenarios of merging BNSs using the population synthesis code COSMIC, for one chosen set of physical parameters (Sect.~\ref{sec:pop-synth}). We discussed the properties of the stellar progenitors at different metallicities, as well as the resulting BNS populations and studied the predicted merger rates. Our results can be summarized as follows:
\begin{itemize}
    \item We identified three dominant evolutionary tracks: the \emph{standard} track, also commonly studied in other works, that involves a phase of CE with the first-born NS, followed by an episode of \textit{case BB} mass transfer; the \emph{equal-mass} track where both progenitor stars evolve on synchronized timescales, and which includes a phase of CE when both stars are in their giant phase; and a track that involves the formation of an O-Ne WD, followed by its AIC due to mass accretion from the companion. 
    \item With our choice of model, we found that the \emph{equal-mass} evolutionary track is the major contributor to the population of merging BNSs at low redshift (higher metallicities), while the \emph{AIC} track is dominant at high redshift (lower metallicities). The \emph{standard} track, on the contrary, is always sub-dominant, as it mostly forms systems with $\Delta t_\mathrm{delay} > t_\mathrm{Hubble}$.
    \item Similarly to previous studies, we find that BNSs that merge within the Hubble time undergo at least one episode of CE. We also find a merger rate at $z=0$ in agreement with the constraints from current LIGO/Virgo observations, although more detections are needed to obtain better constraints. The merger rate peaks at redshift $z\approx 1.7$.
    \item The distribution of time delays between the formation of stellar progenitors and the BNS merger does not follow a simple power-law \emph{at a fixed progenitor metallicity}. Instead, the distribution is typically complex, its shape evolves with metallicity and depends on the relative contribution of the different evolutionary tracks. 
    \item In some cases, particularly at low metallicities, the stellar evolution time (the time to form a BNS) can be longer than the orbital decay time of the BNS due to emission of GWs.
\end{itemize}

There are two aspects of our results which deviate from previous similar studies of BNS populations: \citet{2018MNRAS.481.1908K}, \citet{2018A&A...615A..91B}, \citet{2018MNRAS.481.4009V} and \citet{2023MNRAS.524..426I}, that used the \textsc{ComBinE}, \texttt{StarTrack}, COMPAS and SEVN population synthesis codes, respectively. These studies find that the \emph{standard} track is dominant at all redshifts; and they do not exhibit the \emph{AIC} track. One reason for these discrepancies may be the difference of population synthesis codes used in these studies, as well as different choices of physical parameters. COMPAS, used by \citet{2018MNRAS.481.4009V}, is the closest population synthesis code to COSMIC, since it is also based on BSE. However, \citet{2018MNRAS.481.4009V} focused on Galactic systems without selecting merging binaries, whereas in this work we only discuss systems merging within the age of the Universe. In view of our results, it is plausible that these two distinct populations could have different formation channels, and we plan to further explore this topic in future work. In particular, the \emph{AIC} track that only appears at high redshift, produces very short merger-time systems and may therefore  be exclusively related to merging systems. \citet{2018A&A...615A..91B} focused on merging systems and assumed that all the progenitor stars formed at a single metallicity of $Z=0.01$ in their isolated binary formation scenario. The dominant evolutionary track in their work corresponds to our \emph{standard} track, with two phases of CE owing to an unstable mass transfer from the stripped helium companion with the first-born NS. In this work, we find a dominant contribution of the \emph{equal-mass} track at such a slightly sub-solar metallicity. This difference could be due to a different choice of binary evolution parameters, and will be explored in a follow-up study.\citet{2018MNRAS.474.2937C} show that the \emph{AIC} evolutionary track may contribute to the population of merging BNSs, though it is subdominant at the three metallicities they sample. Similarly, \citet{2012ApJ...759...52D} show a contribution of the \emph{AIC} track of $8\%$ at solar metallicity, which disappears at $0.1~Z_\odot$. These results are also at odds with our findings. As discussed in \citet{2019MNRAS.484..698R}, the properties of BNS systems formed in the \emph{AIC} track may be particularly sensitive to CE physics, but the contribution of this track may be significant. Finally, we stress that COSMIC relies on pre-computed single stellar evolution tracks. Several recent population synthesis codes feature their own evolutionary tracks, most notably POSYDON \citep{2023ApJS..264...45F} and SEVN \citep{2023MNRAS.524..426I}, which are based on more recent results using MESA (\citealt{2023ApJS..265...15J} and references therein) and \textsc{Parsec} \citep{2012MNRAS.427..127B}. They rely on interpolations between the tracks rather than on fitting formulae and lead to substantial differences in the populations of compact objects formed. In light of these results, a dedicated comparison between the various population synthesis codes would be needed in order to fully understand the sources of the discrepancies we mentioned. 

While we did not perform a full parameter study, we looked into different values for the critical mass ratios $q_\mathrm{crit}$, using \texttt{qcflag} (see Sect.~\ref{sec:pop-synth}). We observed  that this flag has a strong effect on binary evolution. When setting \texttt{qcflag}~$ = 1~\mathrm{or}~2$, the mass transfers from stripped helium stars are often dynamically unstable (no \textit{case BB} mass transfer) and lead to a new phase of CE, while in the model presented here (\texttt{qcflag}~$=5$) they are always dynamically stable. Our choice of model therefore naturally produces evolutionary tracks with only one CE phase, while the others have two, even sometimes three CE events.

We also tested another aspect of CE physics. In our model, we assume that stars automatically merge during CE if the companion has no core-envelope boundary (\texttt{cemergeflag}~$=1$). We tested the effect of this assumption by allowing such binaries to survive CE (\texttt{cemergeflag}~$=0$), and found that this typically adds a few additional evolutionary \emph{sequences}. In particular, in this alternative scenario, there is an additional \emph{AIC} evolutionary sequence which dominates at high metallicities, and where a CE appears when the primary has already formed a WD and the secondary is on the HG. These preliminary tests emphasize the importance of a full-scale parameter study that we leave for future work. 

In this work, we did not address the question of NS masses. Indeed, there are only two observed merging BNSs as of writing. The observed Galactic population of NSs gives more insight into their actual mass distribution which, if coupled with another treatment of NS mass in population synthesis codes (as in e.g. COMBINE; \citealt{2018MNRAS.481.1908K}), could provide additional parameter constraints. This interesting aspect should be investigated more in the future.

The results presented in this work, if confirmed, could be important for the study of the r-process abundances in the interstellar medium. Indeed, the multi-messenger observations of GW~170817 have proved that r-process elements are synthesized in the ejecta of merging BNSs, and power kilonova emission \citep{2017ApJ...848L..12A}. It seems natural to assume that all the r-process elements in the Galaxy were forged in BNS mergers, since the overall merger rates are consistent with the estimated total mass in r-process elements, though the predictions for the exact yields per event are still uncertain, owing to the effect of e.g. the equation of state, or the ejecta geometry. However, detailed semi-analytic and hydrodynamical models of the Milky Way show that in order for BNS mergers to be the dominant source of r-process elements in metal-poor stars, the delay time between the formation of progenitor stars and BNS merger $\Delta t_\mathrm{delay}$ should be very short, of the order of a few Myr, or at most tens of Myr \citep[e.g.][]{2018MNRAS.478.1994B,2019MNRAS.487.4847B,2020MNRAS.494.4867V,2023ApJ...943L..12K}. Such delays are much shorter than those predicted for the \emph{standard} evolutionary track. In our study, BNSs formed via the \emph{AIC} track have very short delay times and are also more commonly formed at lower metallicities. Because the AIC events do not eject final fusion products such as iron, unlike core-collapse SNe, this evolutionary track allows the local enrichment of the surrounding gas in r-process elements, with a limited enrichment in nuclear fusion products. This could therefore naturally explain the existence of low-metallicity stars enriched with r-process elements.

Statistical studies of the offsets of short gamma-ray bursts (GRBs) in their host galaxies \citep[e.g.][]{2022ApJ...940...56F} show that a fraction of short GRBs are observed close to the center of their host galaxies. Such observations are indirect evidence that progenitors to short GRBs (often associated to BNS mergers) should have low natal kicks or short merger times. Other studies have pointed out that most short GRB host galaxies are still star-forming \citep{2022ApJ...940...57N}, which again hints towards a population of short delay time BNS mergers. Our results naturally provide such a population, mostly in the \emph{AIC} track.

Finally, our results could be important for the next generation of GW detectors. Indeed, while current interferometers can only observe BNSs in the local Universe, the planned third-generation detectors, Einstein Telescope and Cosmic Explorer, \citep{2020JCAP...03..050M, 2021arXiv210909882E, 2023JCAP...07..068B}, will see much farther out, reaching the cosmic noon and beyond. These observations will therefore give us access to a much wider range of environments of BNS progenitors, and help us uncover their origins.

\begin{acknowledgements}

CP acknowledges funding support from the Initiative Physique des Infinis (IPI), a research training program of the Idex SUPER at Sorbonne Université. This work made use of the Infinity cluster, hosted by the Institut d’Astrophysique de Paris, on which the simulations were run and post-processed. The authors warmly thank S. Rouberol for running it smoothly. The authors thank K. Breivik for useful discussions and the entire COSMIC team for making the code publicly available.

\end{acknowledgements}

\bibliographystyle{aa}
\bibliography{biblio}

\appendix

\section{Parameters used in the COSMIC simulations}\label{ap:cosmic_params}

We list in Table~\ref{tab:cosmic_params} the parameter values used in our simulation (see Sec.~\ref{sec:pop-synth} for a more detailed discussion on our choice of parameters). The version of COSMIC used in this work is v.~3.4.0.

\begin{table}
    \caption{List of the parameter values used in our COSMIC simulations.}
    \centering
    \resizebox{0.73\hsize}{!}{
    \begin{tabular}{c|c}
    \hline\hline
        \textbf{Parameter / Flag} &  \textbf{Value}\\ \hline
        sampling\_method & independent \\
        primary\_model & kroupa01 \\
        porb\_model & sana12 \\
        ecc\_model & sana12 \\
        qmin & $-1$ \\
        binfrac\_model & $0.5$ \\
        metallicity & from $0.000095$ to $0.014$ \\
        seed & $42$ \\
        pts1 & $0.001$ \\
        pts2 & $0.01$ \\
        pts3 & $0.02$ \\
        zsun & $0.014$ \\
        windflag & $3$ \\
        eddlimflag & $0$ \\
        neta & $0.5$ \\
        bwind & $0.0$ \\
        hewind & $0.5$ \\
        beta & $-1$ \\
        xi & $0.5$ \\
        acc2 & $1.5$ \\
        alpha1 & $1.0$ \\
        lambdaf & $0.0$ \\
        ceflag & $1$ \\
        cekickflag & $2$ \\
        cemergeflag & $1$ \\
        cehestarflag & $0$ \\
        qcflag & $5$ \\
        qcrit\_array & default \\
        kickflag & $0$ \\
        sigma & $265.0$ \\
        bhflag & $1$ \\
        bhsigmafrac & $1.0$ \\
        sigmadiv & $-20.0$ \\
        ecsn & $2.25$ \\
        ecsn\_mlow & $1.6$ \\
        aic & $1$ \\
        ussn & $1$ \\
        pisn & $-2$ \\
        polar\_kick\_angle & $90.0$ \\
        natal\_kick\_array & default \\
        remnantflag & $3$ \\
        mxns & $3.0$ \\
        rembar\_massloss & $0.5$ \\
        bhspinflag & $0$ \\
        bhspinmag & $0.0$ \\
        grflag & $1$ \\
        eddfac & $1.0$ \\
        gamma & $-2$ \\
        don\_lim & $-1$ \\
        acc\_lim & $-1$ \\
        tflag & $1$ \\
        ST\_tide & $1$ \\
        fprimc\_array & default \\
        ifflag & $0$ \\
        wdflag & $1$ \\
        epsnov & $0.001$ \\
        bdecayfac & $1$ \\
        bconst & $3000$ \\
        ck & $1000$ \\
        rejuv\_fac & $1.0$ \\
        rejuvflag & $0$ \\
        bhms\_coll\_flag & $0$ \\
        htpmb & $1$ \\
        ST\_cr & $1$ \\ \hline
    \end{tabular}
    }
    \label{tab:cosmic_params}
\end{table}

\section{ZAMS progenitor properties evolution with metallicity} \label{ap:initial_properties}

\begin{figure}[!b]
    \centering
    \resizebox{\hsize}{!}{\includegraphics{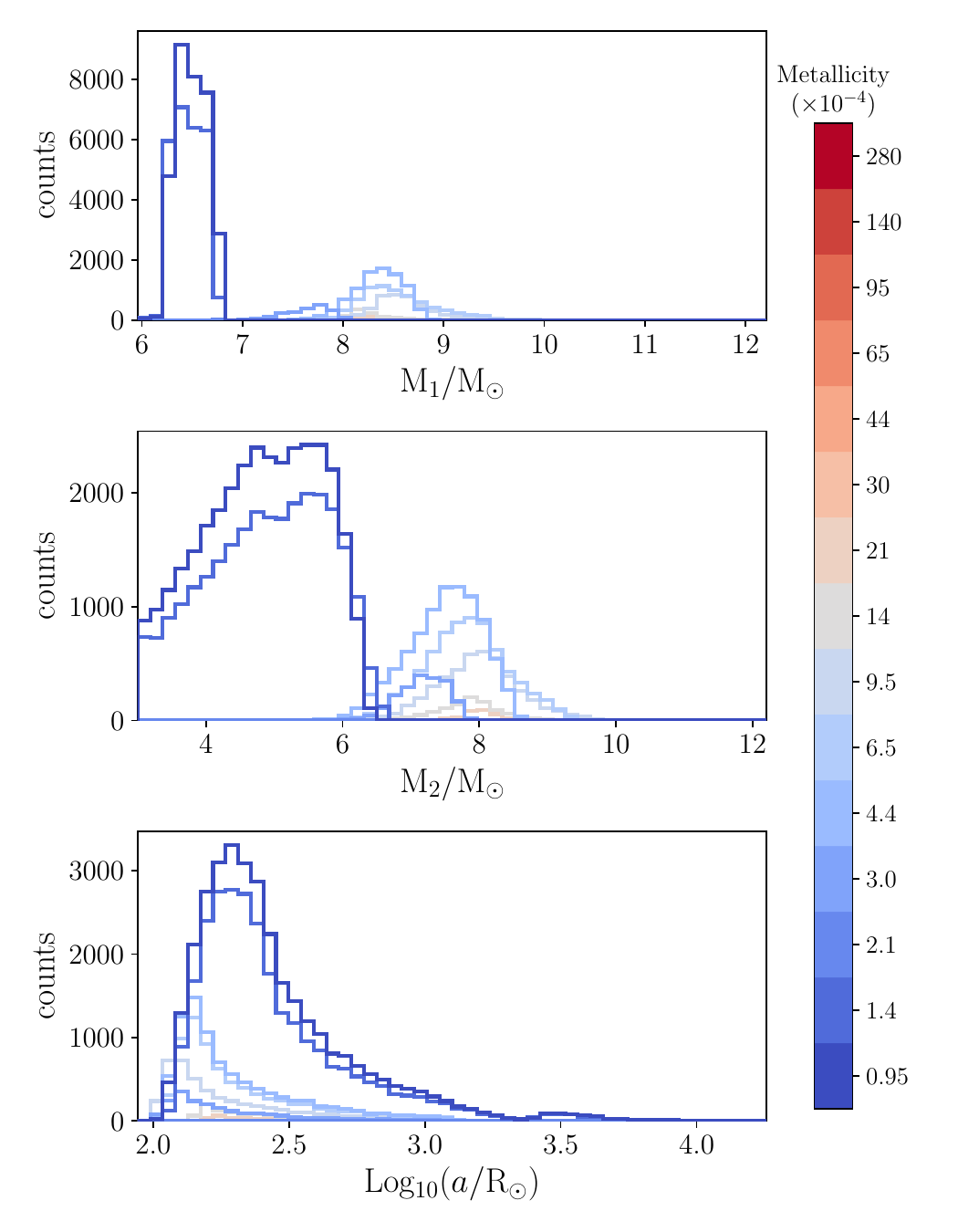}}
    \caption{Distributions of stellar progenitors properties at ZAMS for the \emph{standard} evolutionary track and for various metallicities. From top to bottom, the three panels show the distributions of (1) primary masses; (2) secondary masses; (3) initial semi-major axes.}
    \label{fig:initial_standard}
\end{figure}

\begin{figure}
    \centering
    \resizebox{\hsize}{!}{\includegraphics{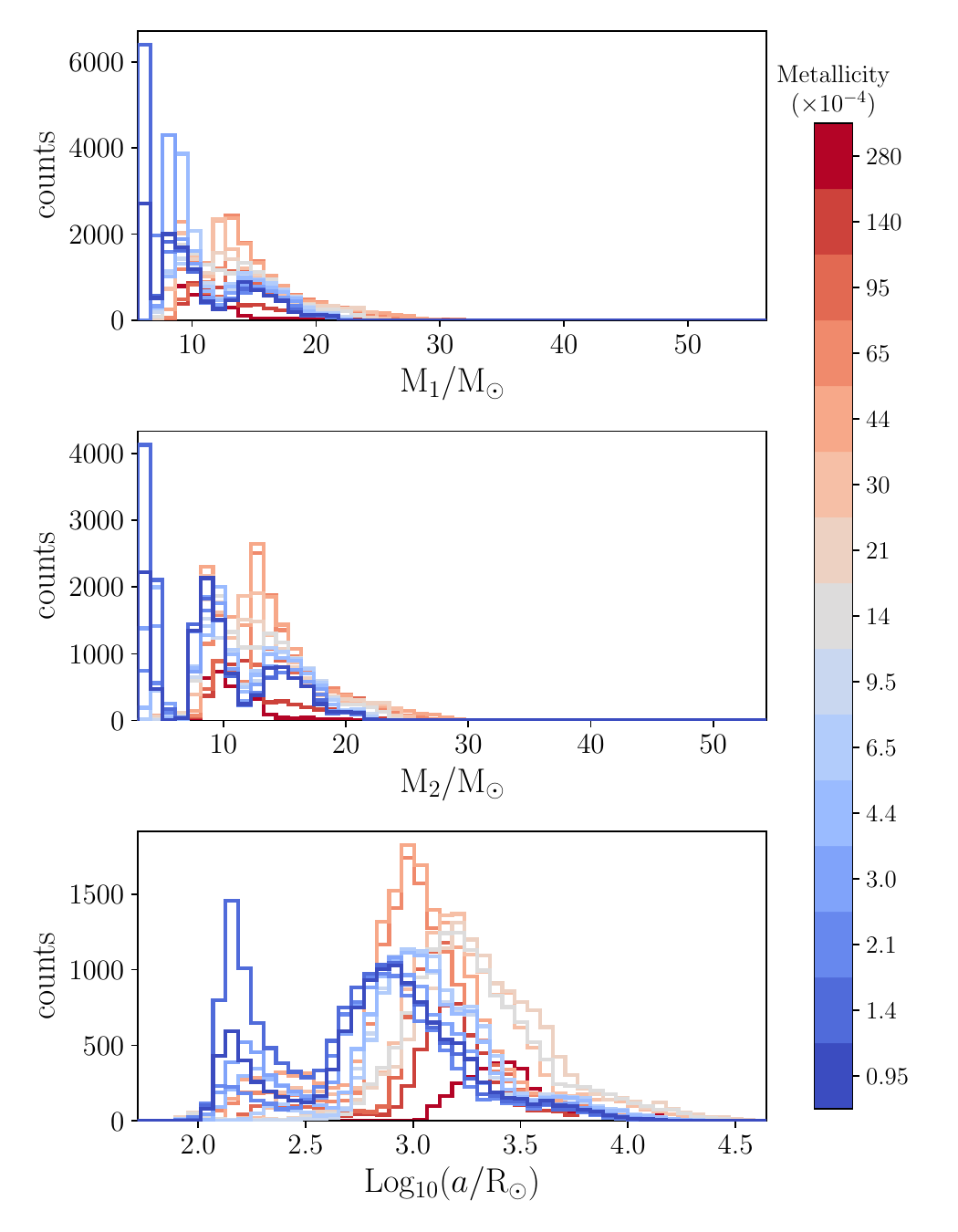}}
    \caption{Same as Fig.~\ref{fig:initial_standard} for the \emph{equal-mass} evolutionary track.}
    \label{fig:initial_equal_mass}
\end{figure}

We show here the distributions of the progenitor properties at ZAMS for the \emph{standard} evolutionary track (Fig.~\ref{fig:initial_standard}) and the \emph{equal-mass} evolutionary track (Fig.~\ref{fig:initial_equal_mass}). These figures are discussed in Sect.~\ref{sec:fixed_met_progenitors}.

\end{document}